\documentclass[10pt, aps, prd, amsmath, amssymb, amsfonts, floatfix, twocolumn, superscriptaddress, nofootinbib, longbibliography]{revtex4-2}
\usepackage{xcolor}
\usepackage{bm}
\usepackage{times}
\usepackage{dcolumn}
\usepackage{enumerate}
\usepackage{graphicx}
\usepackage[utf8]{inputenc}
\usepackage{latexsym}
\usepackage{rotating}
\usepackage{orcidlink}
\usepackage{hyperref}
\usepackage{subcaption}
\usepackage{caption}
\usepackage{booktabs}
\usepackage{multirow}
\usepackage{float}

\hypersetup{
    colorlinks=true,
    linkcolor=blue,
    citecolor=blue,
    urlcolor=blue
}

\begin{document}

\title{Higher Lovelock Curvature Terms Favor Local Nakedness in Dust Collapse}

\author{Apratim Ganguly\orcidlink{0000-0001-7394-0755}}
\email{apratim@iucaa.in}
\affiliation{Inter-University Centre for Astronomy and Astrophysics, Post Bag 4, Ganeshkhind, Pune 411007, India}
\author{Radouane Gannouji\orcidlink{0000-0003-0749-7593}}
\email{radouane.gannouji@pucv.cl}
\affiliation{Instituto de Física, Pontificia Universidad Católica de Valparaíso Av. Brasil 2950, Valparaíso, Chile.}
\author{Akshay Kumar\orcidlink{0000-0003-1478-0223}}
\email{akshay.relativity@gmail.com}
\affiliation{Department of Physics \& Astronomical Science, Central University of Himachal Pradesh, Dharamshala-
176215, India}

\date{\today}

\begin{abstract}
We show that higher-curvature Lovelock terms do not restore local cosmic censorship in spherical dust collapse, but instead promote the local visibility of central shell-focusing singularities. On the collapse branch with positive highest-order Lovelock coefficient \(c_N\), the highest nonvanishing Lovelock order \(N\) controls both the near-singularity collapse and the formation of trapped surfaces. In noncritical dimensions, \(D-1-2N>0\), the apparent-horizon curve approaches the singularity curve with trapping exponent \(\beta_N=(D-1)/(D-1-2N)\). Comparing this scale with the first nonvanishing correction \(r^\ell\) to the singularity curve gives the local-visibility condition \(\ell<\beta_N\), provided the singularity curve opens outward. Thus increasing \(N\) enlarges the class of inhomogeneous initial data producing outgoing radial null rays from the central singularity. In the critical odd-dimensional branch, \(D=2N+1\), no apparent horizon forms sufficiently close to the center, so any outward opening of the singularity curve gives local visibility. The locally visible singularities are Królak-strong along the emerging null rays, with Tipler strength reached at threshold. For bound and unbound collapse, the noncritical exponents are unchanged: the energy function modifies the opening of the singularity curve, while in the critical branch it enters the leading terminal collapse velocity. 
\end{abstract}

\maketitle

\section{Introduction}
\label{sec:intro}

Gravitational collapse provides one of the most direct settings in which the
causal structure of spacetime is tested in the strong-field regime. In general
relativity (GR), the collapse of matter can lead to curvature singularities,
and the cosmic censorship problem asks whether such singularities are
necessarily hidden behind horizons or whether they can be visible to external
observers. The homogeneous dust model of Oppenheimer and Snyder provides the
standard collapse-to-black-hole reference point~\cite{Oppenheimer:1939ue},
while Penrose's formulation of cosmic censorship made the visibility of
collapse singularities a central question~\cite{Penrose:1969pc}. Already in
spherical inhomogeneous dust collapse, described by the
Lemaître--Tolman--Bondi (LTB) framework, the answer depends sensitively on
the initial data: sufficiently inhomogeneous density profiles can produce
outgoing null geodesics from the central shell-focusing singularity, leading
to local nakedness~\cite{Eardley:1978tr,Christodoulou:1984mz,Newman:1985gt,
Joshi:1993zg,Singh:1994tb,Joshi:2008zz}.

It is natural to ask how this picture changes once higher-curvature
corrections are included. Lovelock gravity is the distinguished
higher-dimensional extension of Einstein gravity whose field equations remain
second order in derivatives of the metric~\cite{Lovelock:1971yv,
Zumino:1985dp,Charmousis:2008kc}. It provides a controlled framework in
which higher powers of curvature contribute dynamically without introducing
higher-derivative equations of motion. Since these higher-curvature terms
become important precisely in high-curvature regimes, gravitational collapse
is a natural arena in which to study their effect on horizon formation and on
the local validity of cosmic censorship.

In this work we analyze spherical dust collapse in general Lovelock gravity.
We denote by \(N\) the highest nonvanishing Lovelock order included in the
theory. We work on the collapse branch for which the highest Lovelock
coefficient satisfies \(c_N>0\). This sign choice is important: it is the
branch on which the large-curvature root of the Lovelock polynomial is real
and positive, and it is also the branch on which the highest Lovelock term
produces the delayed or absent near-center trapping derived below.

The simplest nontrivial example is Einstein--Gauss--Bonnet (EGB) gravity.
Previous studies have shown that Gauss--Bonnet corrections can change the
relative timing of singularity and apparent-horizon formation, and that the
five-dimensional case has a causal structure qualitatively different from the
higher-dimensional cases~\cite{Maeda:2006pm,Jhingan:2010zz, Chatterjee:2021zre}. These results
suggest that higher-curvature corrections do not necessarily strengthen cosmic 
censorship. A systematic treatment for general Lovelock order and arbitrary 
spacetime dimension, however, requires identifying which features are special 
to Gauss--Bonnet gravity and which follow from the general Lovelock structure.

Related studies of Lovelock dust collapse, charged Lovelock dust collapse,
pure Lovelock collapse, and third-order Lovelock collapse also show that the
fate and curvature strength of the singularity depend strongly on dimension
and on the Lovelock order~\cite{Ohashi:2011zza,Ohashi:2012wfa,
Dadhich:2013bya,Dialektopoulos:2023qda,Kumar:2025qqj, Brassel:2022mss, Dadhich:2016wtb}. In particular, it was
observed in third-order Lovelock collapse that the critical dimension
\(D=7=2\times3+1\) is qualitatively special, and the corresponding pattern
for general Lovelock order was suggested~\cite{Zhou:2011vz}. One of the
results of the present work is to derive the local version of this pattern
for arbitrary highest Lovelock order \(N\): the critical case \(D=2N+1\)
eliminates apparent horizons sufficiently close to the center, while for
\(D>2N+1\) the central singularity is covered or visible according to a
finite trapping scale.

Our main question is whether higher-curvature Lovelock terms tend to restore
local cosmic censorship or instead enlarge the class of initial data giving
locally visible singularities. The answer is controlled by two universal
exponents. The first is the near-singularity collapse exponent
\begin{equation}
    \alpha_N=\frac{2N}{D-1},
\end{equation}
which follows from the dominance of the highest Lovelock term as the
shell-focusing singularity is approached. The second is the apparent-horizon
delay exponent
\begin{equation}
    \beta_N=\frac{D-1}{D-1-2N},
\end{equation}
defined when \(D-1-2N>0\). This exponent determines how the
apparent-horizon curve approaches the central singularity. The relation
\begin{equation}
    \beta_N=\frac{1}{1-\alpha_N}
\end{equation}
shows that the trapping analysis and the outgoing null-geodesic analysis are
governed by the same local scaling.

For marginally bound collapse, let the singularity curve have the expansion
\begin{equation}
    t_s(r)=t_0+\chi_\ell r^\ell+O(r^{\ell+1}),
\end{equation}
where \(\ell\) is the order of the first nonvanishing correction to the
central singularity time. The leading local-visibility criterion in the
noncritical branch is
\begin{equation}
    \ell<\beta_N, \quad \chi_\ell>0.
\end{equation}
The condition \(\chi_\ell>0\) means that neighboring shells become singular
later than the central shell. Thus increasing the highest Lovelock order
increases the trapping-delay exponent and makes local visibility possible
for a larger class of initial profiles. In the critical case
\(D-1-2N=0\), corresponding to maximal Lovelock order in odd spacetime
dimensions, the highest-order term in the apparent-horizon equation is
independent of the horizon radius. For \(c_N>0\), no apparent horizon forms
sufficiently close to the central singularity, and outgoing null rays can
emerge whenever the singularity curve opens outward.

This criterion also clarifies the smooth-data sector of pure
Gauss--Bonnet collapse. For \(N=2\), a generic smooth density profile has
\(\ell=2\). The noncritical visibility condition then gives locally visible
central singularities in \(D=6,7,8\), while \(D=9\) is the threshold case
and smooth data are locally covered for \(D\geq10\). The critical dimension
\(D=5\), where \(D=2N+1\), is separate: no apparent horizon forms
sufficiently close to the center. This is consistent with recent pure
Gauss--Bonnet analyses reporting covered central singularities for smooth
admissible profiles and, in particular, hidden singularities from the
asymptotic observer under their assumptions~\cite{Dialektopoulos:2023qda,
Kumar:2025qqj}. The comparison should be read with two distinctions in mind:
our result is local rather than global, and our main analysis is
pressureless dust, whereas the recent pure Gauss--Bonnet work also considers
perfect and viscous fluids.

This conclusion is also physically natural \cite{Dadhich:2013bya}. 
In higher-dimensional Einstein gravity the gravitational force becomes more singular at short distances, \(r^{-(D-2)}\), so increasing the dimension tends to strengthen trapping and favor black-hole formation. 
Higher Lovelock terms change this short-distance behavior. 
The effective gravitational field generated by the dominant Lovelock contribution is less singular than the Einstein one, so the collapse can reach the shell-focusing singularity before trapped surfaces have enough time to close around the center. 
In this sense, the tendency toward local nakedness is not an accident of the algebra: it reflects the softening of the near-center gravitational field by higher-curvature Lovelock terms.

The three possible local outcomes are summarized schematically in Fig.~\ref{fig:visibility}. 

\begin{widetext}
\noindent
\begin{minipage}{\linewidth}
\centering
\includegraphics[width=0.9\textwidth]{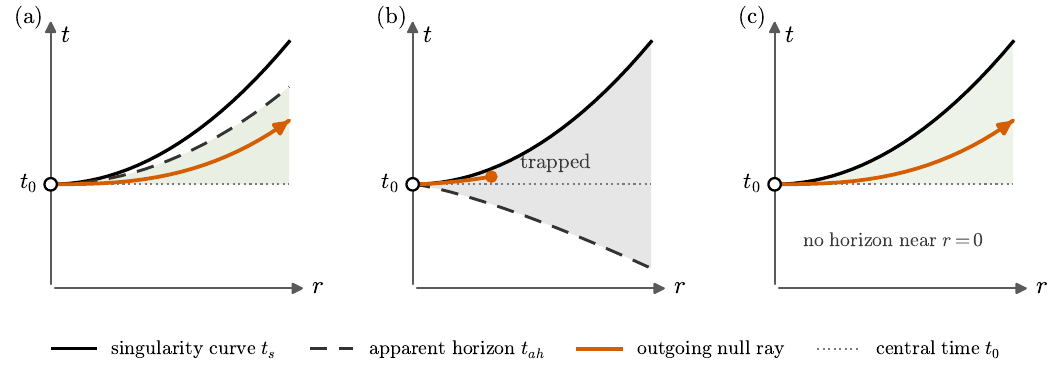}
\captionof{figure}{\small Schematic local causal structure near the central shell-focusing singularity in the comoving \((r,t)\) plane.
The solid curve is the singularity curve \(t_s(r)\), the dashed curve is the apparent-horizon curve \(t_{\rm ah}(r)\), and the orange arrow denotes an outgoing radial null ray.
For \(\ell<\beta_N\), the null ray emerges through a locally untrapped region.
For \(\ell>\beta_N\), trapped surfaces cover a neighborhood of the central singularity.
In the critical Lovelock case \(D=2N+1\), no apparent horizon forms sufficiently close to the center on the \(c_N>0\) branch.
Curve separations are exaggerated for clarity.}
\label{fig:visibility}
\end{minipage}
\end{widetext}

We also examine the curvature strength of the locally visible singularities.
Along the outgoing radial null rays, the Ricci contraction diverges with a
power determined by the same near-center data. In the strict visible branch
the singularities are Królak-strong, while the Tipler Ricci-focusing
condition is reached at the threshold case. Thus the locally visible
singularities found here are not curvature-regular artifacts of the
coordinate system or of the leading-order approximation.

Although the main derivation is first presented for marginally bound
collapse, we also extend the analysis to bound ($E(r)<0$) and unbound ($E(r)>0$) configurations. In the noncritical branch, the energy function does
not change the universal exponents \(\alpha_N\) and \(\beta_N\), but it can
change the opening of the singularity curve. The relevant power in the
visibility criterion is then the first nonvanishing power \(\eta\) in the
full expansion of \(t_s(r)\), after the density and energy contributions are
combined. In the critical branch, the energy function enters the leading
collapse velocity itself. On the standard positive-coupling branch, this
changes the terminal velocity but does not produce a turning point for
initially collapsing data; choices of \(b_0\) that make the terminal velocity
imaginary are already outside the initially collapsing branch that reaches
the center.

The paper is organized as follows. In Sec.~\ref{sec:setup}, we introduce the
Lovelock action, the generalized LTB interior, and the regularity conditions
on the initial data. In Sec.~\ref{sec:dynamics}, we specialize to marginally
bound collapse and derive the singularity curve and the universal
near-singularity scaling. In Sec.~\ref{sec:horizons}, we analyze apparent
horizons and obtain the trapping-delay exponent \(\beta_N\). In
Sec.~\ref{sec:visibility}, we derive the local-visibility criterion and the
curvature-strength behavior along outgoing null rays. In Sec.~\ref{sec:special},
we recover the Einstein and Einstein--Gauss--Bonnet limits as consistency
checks. In Sec.~\ref{sec:nonmarginal}, we extend the discussion to bound and
unbound collapse. We conclude in Sec.~\ref{sec:conclusions}.

\section{Lovelock Framework and Collapse Setup}
\label{sec:setup}

\subsection{Lovelock action and field equations}
\label{subsec:action}

We consider gravitational dynamics in $D$ spacetime dimensions within the framework of Lovelock gravity, which is the most general metric theory whose field equations are second order in derivatives of the metric~\cite{Lovelock:1971yv,Zumino:1985dp}. 
We take the action to be
\begin{equation}
    S = \int d^D x \, \sqrt{-g} \, \sum_{m=1}^{N} a_m \mathcal{L}_m ,
    \label{eq:lovelock_action}
\end{equation}
where $a_m$ are Lovelock coupling constants. We have omitted a possible cosmological-constant term, since it does not play a role in the local near-singularity analysis considered here. The Lovelock densities are defined by
\begin{equation}
    \mathcal{L}_m = \frac{1}{2^m} \delta^{\nu_1 \nu_2 \cdots \nu_{2m}}_{\mu_1 \mu_2 \cdots \mu_{2m}} R^{\mu_1 \mu_2}{}_{\nu_1 \nu_2} \cdots R^{\mu_{2m-1} \mu_{2m}}{}_{\nu_{2m-1} \nu_{2m}} .
    \label{eq:lovelock_density}
\end{equation}
Here $R^{\mu\nu}{}_{\rho\sigma}$ is the Riemann tensor and
$\delta^{\nu_1 \cdots \nu_{2m}}_{\mu_1 \cdots \mu_{2m}}$ is the generalized totally antisymmetric Kronecker delta. The integer $N$ denotes the highest nonvanishing Lovelock order included in the theory, with
\begin{equation}
    N \leq \left\lfloor \frac{D-1}{2} \right\rfloor , \quad a_N\neq 0 .
    \label{eq:lovelock_order}
\end{equation}
Varying the action with respect to the metric gives the Lovelock field equations
\begin{align}
    -\sum_{m=1}^{N}\frac{a_m}{2^{m+1}}\delta^{\mu \mu_1 \cdots \mu_{2m}}_{\nu \nu_1 \cdots \nu_{2m}} R^{\nu_1 \nu_2}_{\mu_1 \mu_2} \cdots R^{\nu_{2m-1} \nu_{2m}}_{\mu_{2m-1} \mu_{2m}} = T^{\mu}_{\nu}.
    \label{eq:field_eq}
\end{align}
We work in units in which the gravitational coupling has been absorbed into the normalization of the Lovelock coefficients and the matter stress tensor. Unless stated otherwise, we assume that the relevant Lovelock coefficients are chosen so that the collapse branch considered below is real and connected continuously to the Einstein branch at low curvature.

In this work, we focus on spherically symmetric spacetimes in
\begin{equation}
    D=n+2
\end{equation}
dimensions, where $n$ is the dimension of the angular sector. This decomposition is convenient for studying dust collapse, since it allows a direct generalization of the LTB framework to higher-dimensional Lovelock gravity.

\subsection{Interior geometry and matter model}
\label{subsec:interior}

We model the collapsing configuration as a spherically symmetric distribution of pressureless dust. 
The matter content is described by
\begin{equation}
    T_{\mu\nu} = \rho(t,r)\,u_\mu u_\nu ,
    \label{eq:dust_tensor}
\end{equation}
where $\rho(t,r)$ is the energy density and $u^\mu$ is the comoving velocity of the dust, normalized as
\begin{equation}
    u^\mu u_\mu=-1 .
\end{equation}
In comoving coordinates, we take
\begin{equation}
    u^\mu=\delta^\mu_t .
\end{equation}

Under the assumption of spherical symmetry, the interior spacetime can be written in generalized LTB form as
\begin{equation}
    ds^2 = -dt^2 + \frac{R'(t,r)^2}{1+E(r)}\,dr^2 + R(t,r)^2\,d\Omega_n^2 .
    \label{eq:ltb_metric}
\end{equation}
Here $R(t,r)$ is the areal radius, a prime denotes differentiation with respect to $r$, and $d\Omega_n^2$ is the metric on the unit $n$-sphere. The function $E(r)$ is the energy function of the shell labeled by $r$. With the convention used in Eq.~\eqref{eq:ltb_metric}, the collapse is classified as
\begin{itemize}
\item $E(r)=0$: marginally bound collapse,
\item $E(r)<0$: bound collapse,
\item $E(r)>0$: unbound collapse.
\end{itemize}
Regularity at the center requires
\begin{equation}
    E(r)=r^2 b(r),
    \label{eq:E_scaling}
\end{equation}
where $b(r)$ is finite as $r\to0$.

It is convenient to introduce the dimensionless collapse variable
\begin{equation}
    R(t,r)=rX(t,r),
    \label{eq:X_def}
\end{equation}
Collapse corresponds to $\dot X<0$, with $X$ decreasing from unity to zero. The condition
\begin{equation}
    X(t,r)=0
\end{equation}
signals the formation of a shell-focusing singularity.

The Lovelock--LTB field equations for inhomogeneous dust collapse take the first-integral form~\cite{Ohashi:2011zza}
\begin{equation}
    \sum_{m=1}^{N} c_m\left(\frac{\dot R^2-E(r)}{R^2}\right)^m = \frac{M(r)}{R^{n+1}} ,
    \label{eq:master_R}
\end{equation}
where $M(r)$ is the mass function of the collapsing cloud. With the normalization of the Lovelock action used in Eq.~\eqref{eq:lovelock_action}, the coefficients $c_m$ are
\begin{align}
    & c_1=a_1,\\
    & c_m = a_m \prod_{p=1}^{2m-2}(n-p) = a_m \frac{(n-1)!}{(n+1-2m)!},
    \quad m\geq2 .
    \label{eq:cm_def}
\end{align}
It is useful to introduce the Lovelock polynomial
\begin{equation}
    P(y) = \sum_{m=1}^{N} c_m y^m,
    \quad y(t,r) = \frac{\dot R^2-E(r)}{R^2}=\frac{\dot X^2-b(r)}{X^2}.
    \label{eq:Lovelock_polynomial}
\end{equation}
In terms of this polynomial, Eq.~\eqref{eq:master_R}
takes the compact form
\begin{equation}
    P(y) = \frac{M(r)}{R^{n+1}}.
    \label{eq:Lovelock_first_integral_polynomial}
\end{equation}
Only the polynomial structure of Eq.~\eqref{eq:master_R} and the highest nonvanishing coefficient $c_N$ will be relevant for the near-singularity analysis. In what follows we take
\begin{equation}
    c_N>0
    \label{eq:cN_positive}
\end{equation}
on the collapse branch under consideration. This condition ensures that the large-\(y\) root of the Lovelock polynomial is real and positive when
\begin{equation}
    P(y) \sim c_Ny^N
\end{equation}
is matched to the positive quantity \(M(r)/R^{n+1}\) near the shell-focusing singularity. It is also the sign choice for which the highest Lovelock term gives the delayed or absent near-center trapping described below.

We define the rescaled mass function $\mu(r)$ by
\begin{equation}
    M(r)=r^{n+1}\mu(r),
    \label{eq:mass_def}
\end{equation}
so that regularity at the center corresponds to finite $\mu(r)$ as $r\to0$. Using Eqs.~\eqref{eq:E_scaling}, \eqref{eq:X_def}, and \eqref{eq:mass_def}, Eq.~\eqref{eq:master_R} becomes
\begin{equation}
    \sum_{m=1}^{N} c_m\left(\frac{\dot X^2-b(r)}{X^2}\right)^m = \frac{\mu(r)}{X^{n+1}} .
    \label{eq:master_X_full}
\end{equation}
This is the basic evolution equation for Lovelock dust collapse. Its polynomial structure encodes the different Lovelock contributions, while the highest nonvanishing Lovelock order $N$ controls the leading behavior in the high-curvature regime near the shell-focusing singularity.

Throughout the analysis, we assume that shell-crossing singularities do not form before the shell-focusing singularity. This requires
\begin{equation}
    R'(t,r)>0,
    \label{eq:no_shell_crossing}
\end{equation}
so that the comoving shell label $r$ remains well-defined during the collapse.

\subsection{Near-center regularity and density profile}
\label{subsec:regularity}

We now specify the regularity assumptions on the initial data near the center. 
The initial hypersurface is chosen such that
\begin{equation}
    R(0,r)=r, \quad X(0,r)=1,
    \label{eq:initial_X}
\end{equation}
and we restrict to initially collapsing configurations,
\begin{equation}
    \dot X(0,r)<0 .
\end{equation}
Regularity at the center requires the areal radius, mass function, and energy function to have the behavior
\begin{equation}
    R(t,r)\sim r, \quad M(r)\sim r^{n+1}, \quad E(r)=O(r^2),
    \quad r\to0 .
\end{equation}
These conditions are precisely the reason for the definitions
\begin{equation}
    M(r)=r^{n+1}\mu(r), \quad E(r)=r^2b(r),
\end{equation}
introduced above. The mass function is assumed to admit a smooth expansion near the center,
\begin{equation}
    \mu(r) = \mu_0+\mu_\ell r^\ell+O(r^{\ell+1}),
    \quad \mu_0>0,
    \quad \ell\geq1,
\label{eq:m_expand}
\end{equation}
where $\ell$ denotes the order of the first nonvanishing correction to the central value.

The density profile is related to the mass function by the $tt$ component of the Lovelock--LTB field equations. With the normalization used in Eq.~\eqref{eq:master_R}, one has
\begin{equation}
    \rho(t,r) = \frac{n}{2}\frac{M'(r)}{R^n R'} .
    \label{eq:rho_general}
\end{equation}
On the initial hypersurface, using $R(0,r)=r$, this gives
\begin{equation}
    \rho(0,r) = \frac{n}{2}\left[(n+1)\mu(r)+r \mu'(r)\right].
    \label{eq:density_relation}
\end{equation}
Equivalently,
\begin{equation}
    \mu(r) = \frac{2}{n r^{n+1}}\int_0^r\rho(0,x)\,x^n\,dx .
    \label{eq:m_from_rho}
\end{equation}
Using Eq.~\eqref{eq:m_expand}, the initial density expands near the center as
\begin{equation}
    \rho(0,r) = \frac{n(n+1)}{2}\mu_0 + \frac{n(n+1+\ell)}{2}\mu_\ell r^\ell +
    O(r^{\ell+1}) .
    \label{eq:rho_expand}
\end{equation}
Since \(\mu_0>0\), the central density is positive,
\begin{equation}
    \rho(0,0) = \frac{n(n+1)}{2}\mu_0 > 0 .
\end{equation}
A negative coefficient \(\mu_\ell\) corresponds to a density decreasing away from the center, but it does not violate positivity near \(r=0\), because the leading positive term proportional to \(\mu_0\) dominates for sufficiently small \(r\). Conversely, \(\mu_\ell>0\) corresponds to an initially increasing density profile near the center.

The initial data are therefore specified by the two functions $\mu(r)$ and $b(r)$, subject to the regularity conditions above. 
The marginally bound case corresponds to
\begin{equation}
    b(r)=0,
\end{equation}
while $b(r)<0$ and $b(r)>0$ describe bound and unbound collapse, respectively. 
In the main analysis we focus on the marginally bound case, for which the singularity curve and the trapping time can be treated analytically. 
The extension to non-marginal collapse is discussed separately in Sec.~\ref{sec:nonmarginal}.

The leading behavior of $\mu(r)$ near the center, encoded in $\mu_0$ and $\mu_\ell$, controls the first correction to the singularity curve. 
This correction will later be compared with the near-center behavior of the apparent horizon in order to determine whether outgoing null geodesics can emerge from the central shell-focusing singularity.

\section{Marginally Bound Collapse and Singularity Formation}
\label{sec:dynamics}

We now specialize the Lovelock--LTB evolution equation to marginally bound collapse and analyze the formation of the central shell-focusing singularity. The marginally bound case is technically simple because the evolution equation can be written as a polynomial equation for a single variable, allowing the singularity curve to be obtained in closed integral form.

\subsection{Marginally bound specialization}
\label{subsec:marginal}

The marginally bound case corresponds to
\begin{equation}
    b(r)=0, \quad E(r)=0 .
\end{equation}
The master equation~\eqref{eq:master_X_full} then reduces to
\begin{equation}
    P(y) = \frac{\mu(r)}{X^{n+1}}\,,\quad \text{with} ~~~P(y)=\sum_{m=1}^{N}
    c_m y^m\,,\quad y= \frac{\dot X^2}{X^2}
    \label{eq:master_X_marginal}
\end{equation}
Throughout the analysis we work on a single positive real branch of the polynomial equation. More precisely, on the interval of $y$ values used below we assume
\begin{equation}
    P(y)>0, \quad P'(y)>0,
\end{equation}
so that $P(y)=\mu/X^{n+1}$ has a unique positive solution $y(X,r)$. This assumption is automatic for positive Lovelock coefficients on the positive-$y$ branch, and for mixed-sign couplings it should be understood as part of the branch choice connected to the Einstein branch. 

We focus on the collapsing branch, $\dot X<0$. Using Eq.~\eqref{eq:master_X_marginal}, this gives $\dot X
=
-X\sqrt{y}$. Since $y$ is implicitly a function of \(X\) and \(r\), the collapse equation can be written as
\begin{equation}
\frac{dX}{dt}
=
-X\sqrt{y(X,r)} .
\label{eq:X_evolution}
\end{equation}
This equation determines the time evolution of each comoving shell.

\subsection{Singularity curve}
\label{subsec:singularity}

A shell-focusing singularity forms when the areal radius of a shell vanishes. Since
\begin{equation}
R(t,r)=rX(t,r),
\end{equation}
this occurs when
\begin{equation}
X(t,r)=0 .
\end{equation}
For each shell labeled by \(r\), we define the singularity time \(t_s(r)\) by
\begin{equation}
X(t_s(r),r)=0 .
\end{equation}
Using Eq.~\eqref{eq:X_evolution}, the time required for a shell to collapse from \(X=1\) to \(X=0\) is
\begin{equation}
t_s(r)
=
\int_0^1
\frac{dX}{X\sqrt{y(X,r)}} .
\label{eq:ts_def_sec3}
\end{equation}
Using Eq.~\eqref{eq:master_X_marginal}, one obtains
\begin{equation}
X
=
\left(
\frac{\mu(r)}{P(y)}
\right)^{\frac{1}{n+1}} .
\label{eq:X_of_y}
\end{equation}
For each comoving shell, i.e. at fixed \(r\), differentiating gives
\begin{equation}
\frac{dX}{X}
=
-\frac{1}{n+1}
\frac{P'(y)}{P(y)}\,dy .
\label{eq:dX_over_X}
\end{equation}
The initial value \(X=1\) corresponds to \(y=y_0(r)\), where
\begin{equation}
P(y_0(r))=\mu(r),
\label{eq:y0_def}
\end{equation}
while the singularity \(X=0\) corresponds to \(y\to\infty\). Therefore Eq.~\eqref{eq:ts_def_sec3} becomes
\begin{equation}
t_s(r)
=
\frac{1}{n+1}
\int_{y_0(r)}^{\infty}
\frac{P'(y)}{P(y)\sqrt y}\,dy .
\label{eq:ts_exact_sec3}
\end{equation}
At the center, \(\mu(r)\to \mu_0\). We define \(y_*\) by
\begin{equation}
P(y_*)=\mu_0 ,
\label{eq:ystar_def}
\end{equation}
and the central singularity time by
\begin{equation}
t_0
\equiv
t_s(0)
=
\frac{1}{n+1}
\int_{y_*}^{\infty}
\frac{P'(y)}{P(y)\sqrt y}\,dy .
\label{eq:t0_def}
\end{equation}
Since \(P(y_0)=\mu\), differentiating the lower limit gives
\begin{equation}
\frac{\partial t_s}{\partial \mu}
=
-\frac{1}{(n+1)\mu\sqrt{y_0}}
<0 .
\end{equation}
Thus increasing the local mass decreases the collapse time on the branch considered here.

To determine the behavior of the singularity curve near the center, we use the expansion \eqref{eq:m_expand}. Expanding Eq.~\eqref{eq:ts_exact_sec3} for small \(r\), one finds
\begin{equation}
t_s(r)
=
t_0+\chi_\ell r^\ell+O(r^{\ell+1}),
\label{eq:ts_expand_sec3}
\end{equation}
with
\begin{equation}
\chi_\ell
=
-\frac{\mu_\ell}{(n+1)\mu_0\sqrt{y_*}} .
\label{eq:chi_result}
\end{equation}
Thus, if \(\mu_\ell<0\), corresponding to a density decreasing away from the center, then
\begin{equation}
\chi_\ell>0,
\end{equation}
and the outer shells become singular after the central shell:
\begin{equation}
t_s(r)>t_0
\quad
(r>0).
\end{equation}
This ordering is necessary for the central singularity to have a chance of being locally visible, since outgoing null rays must propagate through regular shells before those shells themselves become singular.

\subsection{Near-singularity Lovelock scaling}
\label{subsec:scaling}

We now analyze the collapse near the shell-focusing singularity, where \(X\to0\). 
From Eq.~\eqref{eq:Lovelock_polynomial}, the right-hand side diverges as \(X\to0\), so
\begin{equation}
y\to\infty .
\end{equation}
Hence the near-singularity regime is controlled by the large-\(y\) behavior of the Lovelock polynomial \(P(y)\), reflecting the dominance of the highest-curvature Lovelock contribution in the high-curvature regime~\cite{Charmousis:2008kc}. Since \(N\) is the highest nonvanishing Lovelock order, the dominant term at large \(y\) is
\begin{equation}
P(y)
\sim
c_N y^N .
\label{eq:P_asymp}
\end{equation}
Substituting this into Eq.~\eqref{eq:Lovelock_polynomial}, we obtain
\begin{equation}
c_N y^N
\sim
\frac{\mu(r)}{X^{n+1}} .
\label{eq:large_y_master}
\end{equation}
This gives on the collapsing branch,
\begin{equation}
\dot X
\sim
-
\left(
\frac{\mu(r)}{c_N}
\right)^{\frac{1}{2N}}
X^{-\frac{n+1-2N}{2N}} .
\label{eq:Xdot_asymp}
\end{equation}
Integrating this expression near \(X=0\), one finds
\begin{equation}
t_s(r)-t
\sim
\frac{2N}{n+1}
\left(
\frac{c_N}{\mu(r)}
\right)^{\frac{1}{2N}}
X^{\frac{n+1}{2N}} .
\label{eq:ts_minus_t_scaling}
\end{equation}
Equivalently,
\begin{equation}
X(t,r)
\sim
A(r)
\left[
t_s(r)-t
\right]^{\alpha_N},
\label{eq:X_scaling}
\end{equation}
where
\begin{equation}
\alpha_N
=
\frac{2N}{n+1}
=
\frac{2N}{D-1},
\label{eq:alphaN}
\end{equation}
and
\begin{equation}
A(r)
=
\left(
\frac{n+1}{2N}
\right)^{\alpha_N}
\left(
\frac{\mu(r)}{c_N}
\right)^{\frac{1}{n+1}} .
\label{eq:A_of_r}
\end{equation}
The near-singularity behavior of the collapse is universally governed by the highest nonvanishing Lovelock order \(N\). In particular:
\begin{itemize}
\item The scaling exponent $\alpha_N={2N}/{(D-1)}$ depends only on \(N\) and the spacetime dimension \(D\).
\item Lower-order Lovelock terms do not contribute to the leading behavior as \(X\to0\).
\end{itemize}
An immediate consequence is that, away from threshold cases, pure Lovelock collapse and Einstein--Lovelock collapse with the same highest nonvanishing order \(N\) have the same leading local visibility criterion. This result reflects the fact that the curvature grows without bound near the shell-focusing singularity, so that the highest power of curvature present in the Lovelock polynomial dominates the collapse dynamics. As a consequence, the approach to the singularity becomes insensitive to lower-order corrections and exhibits a universal scaling behavior determined solely by the highest Lovelock term.

The scaling behavior~\eqref{eq:X_scaling} will play a crucial role in determining the formation of trapped surfaces. In particular, the exponent \(\alpha_N\) controls how rapidly the areal radius shrinks near the singularity, which in turn affects the relative timing between singularity formation and apparent-horizon formation. This connection will be made explicit in Sec.~\ref{sec:horizons}.

When the highest nontrivial Lovelock order allowed in a given dimension is included, the exponent takes a simple parity-dependent form. In odd spacetime dimensions,
\begin{equation}
\alpha_N=1,
\end{equation}
whereas in even dimensions,
\begin{equation}
\alpha_N=\frac{D-2}{D-1}.
\end{equation}
Thus, for maximal Lovelock order, the collapse variable approaches the singularity linearly in \(t_s(r)-t\) in odd dimensions, while in even dimensions it follows the power law with exponent \((D-2)/(D-1)\).

\section{Apparent Horizons and Trapped Surfaces}
\label{sec:horizons}

We now turn to trapped surfaces and apparent horizons in the marginally bound collapse model analyzed above. 
The singularity curve \(t_s(r)\) determines when each shell reaches \(R=0\), but the causal structure is controlled by the formation of trapped regions. 
The relevant quantity is therefore the relative timing between shell-focusing singularity formation and apparent-horizon formation. 
In spherical symmetry, the apparent horizon is characterized by a simple condition involving the areal radius \(R(t,r)\).

\subsection{Apparent-horizon condition}
\label{subsec:horizon}

The apparent horizon is defined as the outermost marginally trapped surface, where the expansion of outgoing radial null geodesics vanishes. 
In spherical symmetry this condition is equivalent to
\begin{equation}
g^{ab}\partial_aR\,\partial_bR=0,
\label{eq:ah_condition_sec4}
\end{equation}
where \(a,b\) run over the two-dimensional \((t,r)\) sector and \(R(t,r)\) is the areal radius~\cite{Hayward:1993wb,Booth:2005qc}. 
For the metric~\eqref{eq:ltb_metric}, one has
\begin{equation}
g^{ab}\partial_aR\,\partial_bR
=
-\dot R^2
+
\frac{1+E(r)}{R'^2}R'^2
=
-\dot R^2+1+E(r).
\end{equation}
Therefore the apparent-horizon condition is
\begin{equation}
\dot R^2-E(r)=1.
\label{eq:ah_R_minus_E}
\end{equation}
Substituting Eq.~\eqref{eq:ah_R_minus_E} into the Lovelock--LTB master equation~\eqref{eq:master_R}, evaluated at \(R=R_{\rm ah}(r)\), gives
\begin{equation}
\sum_{m=1}^{N}
c_m
\left(
\frac{1}{R_{\rm ah}^2}
\right)^m
=
\frac{M(r)}{R_{\rm ah}^{n+1}} .
\label{eq:ah_master_sec4}
\end{equation}
Multiplying by \(R_{\rm ah}^{n+1}\), the apparent-horizon equation becomes
\begin{equation}
\sum_{m=1}^{N}
c_m
R_{\rm ah}^{\,n+1-2m}
=
M(r).
\label{eq:ah_poly_sec4}
\end{equation}
This equation determines the location of the apparent horizon for each shell. 
It also shows explicitly how the different Lovelock orders enter the horizon condition.

Near the center, \(M(r)\to0\). 
The leading behavior of Eq.~\eqref{eq:ah_poly_sec4} is controlled by the highest nonvanishing Lovelock order \(N\). 
Defining
\begin{equation}
s
\equiv 
n+1-2N
=
D-1-2N,
\label{eq:s_def}
\end{equation}
the leading term is
\begin{equation}
M(r)
\sim
c_N R_{\rm ah}^{\,s}.
\label{eq:ah_leading}
\end{equation}
For \(s>0\), this gives
\begin{equation}
R_{\rm ah}(r)
\sim
\left(
\frac{M(r)}{c_N}
\right)^{1/s}.
\label{eq:Rah_scaling_sec4}
\end{equation}
Using \(M(r)=r^{n+1}\mu(r)\), with \(\mu(r)\to \mu_0\), one obtains
\begin{equation}
R_{\rm ah}(r)
\sim
\left(
\frac{\mu_0}{c_N}
\right)^{1/s}
r^{\frac{n+1}{s}}
=
\left(
\frac{\mu_0}{c_N}
\right)^{1/(D-1-2N)}
r^{\frac{D-1}{D-1-2N}} .
\label{eq:Rah_center_scaling}
\end{equation}
Thus, when \(s>0\), the apparent horizon reaches the center with a power determined only by \(D\) and the highest Lovelock order \(N\).

The case \(s=0\), corresponding to odd spacetime dimensions \(D=2N+1\), is qualitatively
different. Then the highest Lovelock contribution to
Eq.~\eqref{eq:ah_poly_sec4} is constant
\begin{equation}
c_N R_{\rm ah}^{0}=c_N .
\end{equation}
Since \(M(r)\to0\) as \(r\to0\), the equation cannot be satisfied by a
solution with \(R_{\rm ah}\to0\) on the branch with \(c_N>0\). Thus, in
the critical Lovelock dimension \(D=2N+1\), no apparent horizon reaches
the center. Therefore, for maximal Lovelock order in odd dimensions, the central singularity is not locally covered by trapped surfaces. 
\subsection{Relative timing of trapping and singularity formation}
\label{subsec:timing}

We now compare the singularity time \(t_s(r)\) with the apparent-horizon time \(t_{\rm ah}(r)\). 
The latter is defined as the time at which the shell labeled by \(r\) satisfies Eq.~\eqref{eq:ah_poly_sec4}. 
In the marginally bound case, \(E(r)=0\), the apparent-horizon condition \eqref{eq:ah_R_minus_E} gives $\dot R^2=1$. Since \(R=rX\), this implies
\begin{equation}
y_{\rm ah}
=
\frac{\dot X^2}{X^2}
=
\frac{1}{R_{\rm ah}^2}.
\label{eq:y_ah}
\end{equation}
Using the marginally bound evolution equation~\eqref{eq:X_evolution}, the apparent-horizon time is
\begin{equation}
t_{\rm ah}(r)
=
\int_{X_{\rm ah}(r)}^1
\frac{dX}{X\sqrt{y(X,r)}} .
\label{eq:tah_def}
\end{equation}
Changing variables from \(X\) to \(y\), as in Sec.~\ref{subsec:singularity}, gives
\begin{equation}
t_{\rm ah}(r)
=
\frac{1}{n+1}
\int_{y_0(r)}^{y_{\rm ah}(r)}
\frac{P'(y)}{P(y)\sqrt y}\,dy .
\label{eq:tah_exact}
\end{equation}
Subtracting this from Eq.~\eqref{eq:ts_exact_sec3}, one obtains
\begin{equation}
t_s(r)-t_{\rm ah}(r)
=
\frac{1}{n+1}
\int_{y_{\rm ah}(r)}^\infty
\frac{P'(y)}{P(y)\sqrt y}\,dy .
\label{eq:time_diff_exact}
\end{equation}
Near the center, \(R_{\rm ah}(r)\to0\) when \(s>0\), and therefore \(y_{\rm ah}\to\infty\). 
The integral is then controlled by the large-\(y\) behavior
\begin{equation}
P(y)\sim c_Ny^N,
\quad
P'(y)\sim Nc_Ny^{N-1}.
\end{equation}
Hence
\begin{equation}
\frac{P'(y)}{P(y)\sqrt y}
\sim
\frac{N}{y^{3/2}} .
\end{equation}
Substitution into Eq.~\eqref{eq:time_diff_exact} gives
\begin{equation}
t_s(r)-t_{\rm ah}(r)
\sim
\frac{2N}{n+1}
y_{\rm ah}^{-1/2}.
\end{equation}
Using Eq.~\eqref{eq:y_ah}, this becomes
\begin{equation}
t_s(r)-t_{\rm ah}(r)
\sim
\frac{2N}{n+1}
R_{\rm ah}(r).
\label{eq:time_diff_R}
\end{equation}
Together with Eq.~\eqref{eq:Rah_center_scaling}, this yields
\begin{equation}
t_s(r)-t_{\rm ah}(r)
\sim
r^{\beta_N},
\label{eq:beta_scaling_sec4}
\end{equation}
where
\begin{equation}
\beta_N
=
\frac{D-1}{D-1-2N},
\label{eq:betaN}
\end{equation}
whenever \(D-1-2N>0\). This exponent depends only on the spacetime dimension \(D\) and on the highest nonvanishing Lovelock order \(N\). It determines how rapidly the apparent-horizon curve approaches the center relative to the singularity curve. Larger values of \(\beta_N\) correspond to a stronger delay of trapping relative to the formation of the central singularity. Within the branch \(D-1-2N>0\), increasing \(N\) at fixed dimension increases \(\beta_N\), and therefore delays the onset of trapped surfaces.

The scaling~\eqref{eq:beta_scaling_sec4} provides the key input for the local-visibility analysis. It must be compared with the first correction to the singularity curve,
\begin{equation}
t_s(r)=t_0+\chi_\ell r^\ell+O(r^{\ell+1}),
\end{equation}
where \(\ell\) characterizes the leading inhomogeneity of the initial density profile. Thus the competition between \(\ell\) and \(\beta_N\) determines whether outgoing null geodesics can emerge from the central shell-focusing singularity before the neighboring shells are trapped. 

The limiting case \(D-1-2N=0\), possible only for maximal Lovelock order in odd spacetime dimensions, is qualitatively different: the highest-order term in the apparent-horizon equation is independent of \(R_{\rm ah}\), and no apparent horizon reaches the center on the \(c_N>0\) branch.

\section{Local Visibility of the Central Singularity}
\label{sec:visibility}

We now determine the conditions under which the central shell-focusing singularity is locally visible. 
The relevant criterion is the existence of future-directed outgoing radial null geodesics with past endpoint at the central singularity. 
Such geodesics must emerge into the regular region of the spacetime and, when trapped surfaces form near the center, they must do so before the neighboring shells become trapped. The analysis below establishes local visibility only. Whether such a null ray reaches the surface of the cloud, or future null infinity after matching to an exterior spacetime, is a separate global question.

\subsection{Outgoing radial null geodesics}
\label{subsec:null}

Radial null geodesics in the spacetime~\eqref{eq:ltb_metric} are obtained by setting \(ds^2=0\) with vanishing angular components. 
For the outgoing branch this gives
\begin{equation}
\frac{dt}{dr}
=
\frac{R'(t,r)}{\sqrt{1+E(r)}} .
\label{eq:null_general}
\end{equation}
In the marginally bound case, \(E(r)=0\), this reduces to
\begin{equation}
\frac{dt}{dr}
=
R'(t,r).
\label{eq:null_eq_sec5}
\end{equation}
Using \(R(t,r)=rX(t,r)\), one has
\begin{equation}
\frac{dt}{dr}
=
X+r\,\partial_r X .
\label{eq:null_X}
\end{equation}
The central shell-focusing singularity is located at \((t,r)=(t_0,0)\). 
It is locally visible if there exists at least one future-directed outgoing radial null geodesic \(t(r)\) such that
\begin{equation}
t(0)=t_0,
\end{equation}
and, for sufficiently small \(r>0\),
\begin{equation}
t(r)<t_s(r),
\quad
t(r)<t_{\rm ah}(r),
\label{eq:visibility_conditions}
\end{equation}
whenever an apparent horizon exists near the center. 
The first inequality ensures that the null ray propagates through a regular shell before that shell becomes singular. 
The second inequality ensures that the ray is not inside the trapped region. 
Thus the question of local visibility is reduced to comparing outgoing radial null geodesics with the singularity curve and the apparent-horizon curve.

\subsection{Near-center analysis}
\label{subsec:nearcenter_visibility}

We analyze Eq.~\eqref{eq:null_X} near the central singularity. 
The singularity curve obtained in Sec.~\ref{subsec:singularity} has the expansion
\begin{equation}
t_s(r)
=
t_0+\chi_\ell r^\ell+O(r^{\ell+1}),
\quad
\ell\geq1 .
\label{eq:ts_expand_visibility}
\end{equation}
We look for an outgoing radial null geodesic emerging from the central singularity in the form
\begin{equation}
t(r)
=
t_0+xr^\gamma+\cdots,
\quad
x>0,
\quad
\gamma>0 .
\label{eq:null_ansatz_sec5}
\end{equation}
The difference between the singularity curve and the null ray is
\begin{equation}
\Delta(r)
\equiv
t_s(r)-t(r).
\label{eq:Delta_def}
\end{equation}
Using Eqs.~\eqref{eq:ts_expand_visibility} and \eqref{eq:null_ansatz_sec5}, this becomes
\begin{equation}
\Delta(r)
=
\chi_\ell r^\ell
-
x r^\gamma
+\cdots .
\label{eq:Delta_expand}
\end{equation}
A necessary condition for the null ray to lie in the regular region is \(\Delta(r)>0\).

Near the shell-focusing singularity, the collapse variable obeys the scaling law derived in Sec.~\ref{subsec:scaling},
\begin{equation}
X(t,r)
\sim
A(r)\,[t_s(r)-t]^{\alpha_N},
\quad
\alpha_N=\frac{2N}{D-1}.
\label{eq:X_scaling_sec5}
\end{equation}
At the center, \(A(r)\to A_0>0\). 
Along the null ray, this gives
\begin{equation}
X(t(r),r)
\sim
A_0\,\Delta(r)^{\alpha_N}.
\label{eq:X_null_scaling}
\end{equation}
We first consider the case in which the singularity curve gives the leading contribution to \(\Delta(r)\), namely
\begin{equation}
\gamma>\ell .
\end{equation}
Then
\begin{equation}
\Delta(r)
\sim
\chi_\ell r^\ell .
\end{equation}
This requires \(\chi_\ell>0\), which means that the neighboring shells become singular after the central shell.

The \(A'(r)\) term in \(R'\) scales as
\begin{equation}
rA'(r)\Delta^{\alpha_N}
=
O(r^{\alpha_N\ell+\ell}) .
\end{equation}
This is down by a factor \(r^\ell\) relative to the leading \(O(r^{\alpha_N\ell})\) terms in \(R'\). Hence only the derivative of the singularity curve contributes at leading order.

Substituting Eq.~\eqref{eq:X_null_scaling} into Eq.~\eqref{eq:null_X}, and taking the derivative \(R'\) at fixed \(t\), one finds the leading behavior
\begin{equation}
R'(t(r),r)
\sim
(1+\alpha_N\ell)
A_0\chi_\ell^{\alpha_N}
r^{\alpha_N\ell}.
\label{eq:null_rhs}
\end{equation}
This expression is also the local shell-crossing check in the visible branch. For $\chi_\ell>0$, $A_0>0$, and $1+\alpha_N\ell>0$, one has $R'>0$ along the emerging null rays and in a sufficiently small neighboring regular region. Thus the local construction is not spoiled by shell crossing near the center.

On the other hand, differentiating Eq.~\eqref{eq:null_ansatz_sec5} gives
\begin{equation}
\frac{dt}{dr}
=
x\gamma r^{\gamma-1}.
\label{eq:null_lhs}
\end{equation}
Matching the leading powers of \(r\) in the null equation gives
\begin{equation}
\gamma
=
1+\alpha_N\ell .
\label{eq:gamma_result}
\end{equation}
The corresponding leading coefficient is
\begin{equation}
x
=
A_0\chi_\ell^{\alpha_N}.
\label{eq:x_result}
\end{equation}

The consistency condition \(\gamma>\ell\) therefore becomes
\begin{equation}
1+\alpha_N\ell>\ell .
\end{equation}
For \(D-1-2N>0\), this is equivalent to
\begin{equation}
\ell
<
\frac{1}{1-\alpha_N}
=
\frac{D-1}{D-1-2N}
=
\beta_N .
\label{eq:beta_condition_sec5}
\end{equation}
Thus the same exponent \(\beta_N\) that controls the apparent-horizon delay also controls the existence of outgoing radial null geodesics in the regular region.

This result has a direct interpretation. 
If \(\ell<\beta_N\), then the separation between the singularity curve and the outgoing null ray is controlled by \(r^\ell\), while the trapping delay behaves as \(r^{\beta_N}\). 
Since \(r^\ell\) dominates over \(r^{\beta_N}\) near the center, the outgoing null ray has enough room to emerge before the neighboring shells are trapped. 
If \(\ell>\beta_N\), the apparent horizon approaches the center too rapidly, and the null ray cannot remain in the untrapped regular region.

The threshold case
\begin{equation}
\ell=\beta_N
\end{equation}
is marginal. 
In that case the null ray, the singularity curve, and the apparent-horizon curve all contribute at the same order, and the visibility depends on the leading coefficients and on subleading terms in the initial data.

\subsection{Visibility criteria}
\label{subsec:criteria}

We now summarize the local visibility criterion for marginally bound Lovelock dust collapse. 
The initial density profile is characterized by
\begin{equation}
\mu(r)
=
\mu_0+\mu_\ell r^\ell+O(r^{\ell+1}),
\quad
\mu_0>0,
\end{equation}
and the singularity curve by
\begin{equation}
t_s(r)
=
t_0+\chi_\ell r^\ell+O(r^{\ell+1}).
\end{equation}
The condition \(\chi_\ell>0\) is necessary for local visibility: it ensures that the central shell becomes singular before the neighboring shells.

When
\begin{equation}
D-1-2N>0,
\end{equation}
the apparent horizon forms near the center and the relevant comparison is between \(\ell\) and
\begin{equation}
\beta_N
=
\frac{D-1}{D-1-2N}.
\end{equation}
At leading order, the result is
\begin{equation}
\begin{cases}
\text{locally visible},
&
\ell<\beta_N
\quad\text{and}\quad
\chi_\ell>0,
\\[5pt]
\text{threshold case},
&
\ell=\beta_N,
\\[5pt]
\text{covered},
&
\ell>\beta_N .
\end{cases}
\label{eq:visibility_summary}
\end{equation}
For \(\ell<\beta_N\), the singularity curve opens sufficiently fast near the center for outgoing radial null geodesics to emerge before trapping. 
For \(\ell>\beta_N\), the apparent horizon reaches the center too early, and the central singularity is covered. 
The case \(\ell=\beta_N\) requires a coefficient-level analysis, because the singularity curve, null ray, and apparent-horizon curve all compete at the same order.

In the critical case
\begin{equation}
D-1-2N=0,
\end{equation}
the apparent horizon does not form in a neighborhood of the center. 
Moreover $\alpha_N=1$ so Eq.~\eqref{eq:gamma_result} gives
\begin{equation}
\gamma=1+\ell>\ell .
\end{equation}
Thus, if \(\chi_\ell>0\), outgoing radial null geodesics can emerge from the central singularity into an untrapped neighborhood. 
For maximal Lovelock order in odd spacetime dimensions, the central singularity is therefore locally visible for density profiles whose first nonvanishing correction gives \(\chi_\ell>0\).

The criterion above shows explicitly how the outcome is controlled by two independent ingredients. 
The initial density profile determines the opening of the singularity curve through \(\ell\) and \(\chi_\ell\), while the Lovelock dynamics determines the trapping delay through \(\beta_N\). 
Increasing the highest Lovelock order increases \(\beta_N\) whenever \(D-1-2N>0\), thereby enlarging the range of initial density profiles for which the central singularity is locally visible.

The parameter \(\ell\) should also be interpreted in relation to the
regularity of the initial density profile. A spherically symmetric scalar
profile that is smooth at the center has vanishing first radial derivative.
Thus the generic smooth case has
\[
\ell=2,
\]
unless the quadratic coefficient is tuned to vanish. By contrast,
\(\ell=1\) corresponds to a nonzero central radial gradient and is not
smooth as a scalar profile on the local Cartesian spatial geometry.

For generic smooth data, the noncritical visibility condition becomes
\[
2<\beta_N
\quad\Longleftrightarrow\quad
N>\frac{D-1}{4}.
\]
Thus, at fixed spacetime dimension, increasing the highest Lovelock order
moves smooth data toward the locally visible regime. Equivalently, for a
given Lovelock order \(N\), generic smooth data are locally visible in the
noncritical branch only below the threshold \(D=4N+1\), with equality giving
the marginal case.

The resulting dimension-by-dimension classification is shown in
Table~\ref{tab:visibility}, while Fig.~\ref{fig:visibility_plane}
displays the same criterion in the \((D,N)\) plane. It shows that, away from the threshold
case \(\ell=\beta_N\), the leading local-visibility criterion depends on
the Lovelock theory only through the highest nonvanishing order \(N\) and
the dimension \(D\). Lower-order Lovelock couplings affect nonuniversal
quantities, such as the central collapse time and the numerical
coefficients in the near-center expansion, but they do not change the
exponents \(\alpha_N\), \(\beta_N\), or the inequality
\(\ell<\beta_N\). Thus pure Lovelock gravity and Einstein--Lovelock gravity
with the same highest order \(N\) give the same leading local-visibility
criterion away from threshold.

\begin{table}[!htbp]
\centering
\caption{\small Local visibility of the central shell-focusing singularity for marginally bound Lovelock dust collapse on the \(c_N>0\) branch. 
Here \(s=D-1-2N\), \(\alpha_N=2N/(D-1)\), and \(\beta_N=(D-1)/(D-1-2N)\). 
The last column gives the outcome for generic smooth data, \(\ell=2\), assuming \(\chi_\ell>0\). 
Critical rows, \(s=0\), have no apparent horizon near the center.}
\label{tab:visibility}
\setlength{\tabcolsep}{8pt}
\begin{tabular}{cccccl}
\toprule
$D$ & $N$ & $s$ & $\alpha_N$ & $\beta_N$ & smooth data ($\ell=2$)\\
\midrule
4 & 1 & 1 & $2/3$ & $3$ & locally visible\\
\midrule
\multirow{2}{*}{5} & 1 & 2 & $1/2$ & $2$   & threshold\\
                   & 2 & 0 & $1$   & ---   & visible (no near-centre AH)\\
\midrule
\multirow{2}{*}{6} & 1 & 3 & $2/5$ & $5/3$ & covered\\
                   & 2 & 1 & $4/5$ & $5$   & locally visible\\
\midrule
\multirow{3}{*}{7} & 1 & 4 & $1/3$ & $3/2$ & covered\\
                   & 2 & 2 & $2/3$ & $3$   & locally visible\\
                   & 3 & 0 & $1$   & ---   & visible (no near-centre AH)\\
\midrule
\multirow{3}{*}{8} & 1 & 5 & $2/7$ & $7/5$ & covered\\
                   & 2 & 3 & $4/7$ & $7/3$ & locally visible\\
                   & 3 & 1 & $6/7$ & $7$   & locally visible\\
\midrule
\multirow{4}{*}{9} & 1 & 6 & $1/4$ & $4/3$ & covered\\
                   & 2 & 4 & $1/2$ & $2$   & threshold\\
                   & 3 & 2 & $3/4$ & $4$   & locally visible\\
                   & 4 & 0 & $1$   & ---   & visible (no near-centre AH)\\
\bottomrule
\end{tabular}
\end{table}

\noindent
\begin{minipage}{\linewidth}
\centering
\includegraphics[width=.95\textwidth]{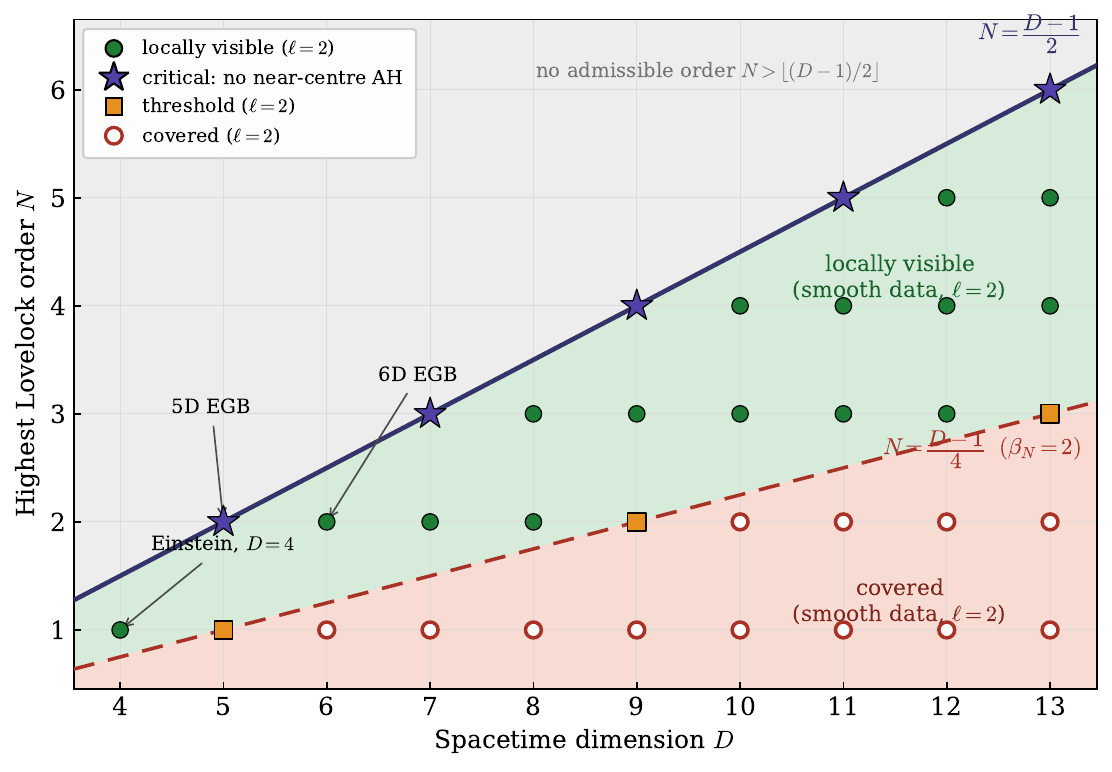}
\captionof{figure}{\small Local visibility of the central singularity in marginally bound Lovelock
dust collapse on the $c_N>0$ branch, in the plane of spacetime dimension $D$ and
highest Lovelock order $N$. Admissible orders satisfy $N\le\lfloor(D-1)/2\rfloor$
(solid line $N=(D-1)/2$); the integer points on this line, realized only in odd
$D$, are critical ($s=D-1-2N=0$), where no apparent horizon forms near the centre
(stars). The dashed line $N=(D-1)/4$ is the smooth-data threshold $\beta_N=2$.
For generic smooth data ($\ell=2$, $\chi_\ell>0$) the central singularity is
locally visible above the dashed line (filled circles) and covered below it (open
circles), with threshold behaviour on it (squares, $D=4N+1$). At fixed $D$,
increasing $N$ moves upward across the threshold: higher Lovelock order promotes
local visibility. Einstein gravity ($N=1$) and Einstein--Gauss--Bonnet gravity
($N=2$) are the lowest two rows.}
\label{fig:visibility_plane}
\end{minipage}
\vspace{-0.5\baselineskip}

\subsection{Curvature strength along the outgoing null ray}
\label{subsec:curvature_strength}

We finally examine the curvature strength of the locally outgoing radial null rays. 
Let \(k^\mu=dx^\mu/d\lambda\) be an affinely parametrized tangent vector, with \(\lambda=0\) at the central singularity. 
Following the standard Tipler--Królak characterization of strong curvature singularities, we consider the Ricci contraction along the null ray,
\begin{equation}
\Psi
\equiv
R_{\mu\nu}k^\mu k^\nu .
\label{eq:Psi_def}
\end{equation}
The behavior of \(\Psi\) and of its affine-parameter integrals distinguishes between Tipler and Królak curvature strength~\cite{Tipler:1977zza,Clarke:1985,Joshi:1996qc}. We use the Ricci contraction \(R_{\mu\nu}k^\mu k^\nu\) because it is the
focusing term entering the null Raychaudhuri equation. It gives a direct
measure of the accumulated convergence of the outgoing null congruence.
This is a Ricci-focusing strength test; a full tidal-strength analysis would
require the complete Riemann tensor, including the Weyl part.

For the marginally bound metric, the radial null condition gives
\begin{equation}
k^t=R'k^r .
\end{equation}
The \(t\)-component of the affine geodesic equation implies
\begin{equation}
\frac{d\ln k^t}{dr}
=
-\dot R' .
\label{eq:kt_geodesic}
\end{equation}
Along the locally visible branch, \(\Delta(r)=t_s(r)-t(r)\sim \chi_\ell r^\ell\), with \(\chi_\ell>0\), and
\begin{equation}
R(t,r)
\sim
rA_0\Delta(r)^{\alpha_N}.
\end{equation}
At fixed \(t\), this gives
\begin{equation}
R'
\sim
(1+\alpha_N\ell)
A_0\chi_\ell^{\alpha_N}
r^{\alpha_N\ell},
\label{eq:Rprime_strength}
\end{equation}
and
\begin{equation}
\dot R'
\sim
-\alpha_N
A_0
\chi_\ell^{\alpha_N-1}
\left[
1-\ell(1-\alpha_N)
\right]
r^{-\ell(1-\alpha_N)} .
\label{eq:Rdotprime_strength}
\end{equation}
For the strict visible branch,
\begin{equation}
\ell<\beta_N
=
\frac{1}{1-\alpha_N},
\end{equation}
the exponent in Eq.~\eqref{eq:Rdotprime_strength} is smaller than unity. Therefore Eq.~\eqref{eq:kt_geodesic} implies that \(k^t\) approaches a finite nonzero limit,
\begin{equation}
k^t\to k^t_0 .
\end{equation}
Since \(d\lambda/dr=R'/k^t\), Eq.~\eqref{eq:Rprime_strength} gives
\begin{equation}
\lambda
\sim
\frac{A_0\chi_\ell^{\alpha_N}}{k^t_0}
r^{1+\alpha_N\ell}.
\label{eq:lambda_r_strength}
\end{equation}

The Ricci contraction can be evaluated geometrically. For the warped product metric,
\begin{equation}
R_{\mu\nu}k^\mu k^\nu
=
-\frac{n}{R}
k^a k^b\nabla_a\nabla_b R ,
\end{equation}
where \(a,b\) refer to the \((t,r)\) sector. Using the radial null condition, this becomes
\begin{equation}
\Psi
=
-\frac{n(k^t)^2}{R}
\left(
\ddot R
-
\frac{\dot R\,\dot R'}{R'}
\right).
\label{eq:Psi_R}
\end{equation}
Keeping the connection terms in \(k^a k^b\nabla_a\nabla_bR\), and using the radial null geodesic equation to eliminate derivatives of \(k^t\), the leading noncancelling term gives
\begin{equation}
\Psi
\sim
\frac{
n\alpha_N(k^t_0)^2
}{
1+\alpha_N\ell
}
\frac{1}{\Delta(r)^2}
\sim
C_\Psi r^{-2\ell},
\label{eq:Psi_r_scaling}
\end{equation}
with \(C_\Psi>0\) on the outgoing branch. The point needed below is the \(\Delta^{-2}\) divergence; the explicit positive coefficient follows only after the Christoffel terms in Eq.~\eqref{eq:Psi_R} are retained. Using Eq.~\eqref{eq:lambda_r_strength}, this can be written as
\begin{equation}
\Psi
\sim
\tilde C_\Psi
\lambda^{-q},
\quad
q
=
\frac{2\ell}{1+\alpha_N\ell},
\quad
\tilde C_\Psi>0 .
\label{eq:Psi_lambda_scaling}
\end{equation}

For \(D-1-2N>0\) and \(\ell<\beta_N\), one has
\begin{equation}
1<q<2 .
\end{equation}
Therefore
\begin{equation}
\int^\lambda \Psi(\lambda')\,d\lambda'
\end{equation}
diverges as \(\lambda\to0\), while the double integral
\begin{equation}
\int^\lambda d\lambda'
\int^{\lambda'} \Psi(\lambda'')\,d\lambda''
\end{equation}
remains finite. Thus the Ricci-focusing integral has the Królak divergence along the outgoing radial null ray, while the corresponding Tipler double integral remains finite in the strict branch $\ell<\beta_N$.

The threshold case \(\ell=\beta_N\) is different. In that case,
\begin{equation}
q=2,
\end{equation}
so a locally visible threshold ray, when it exists, satisfies the corresponding Tipler sufficient curvature-growth condition. In the critical Lovelock case \(D-1-2N=0\), one has \(\alpha_N=1\), and therefore
\begin{equation}
q
=
\frac{2\ell}{1+\ell}.
\end{equation}
For every finite \(\ell\geq1\), this gives \(1\leq q<2\). Hence the critical case is also Królak-strong along the outgoing radial null ray, but it does not satisfy the Tipler Ricci-focusing condition for finite \(\ell\).

Thus the locally visible singularities found above are not curvature-regular artifacts: the Ricci focusing diverges along the outgoing null rays. Their curvature strength, however, is generically of Królak type in the strict visible branch, while Tipler strength arises only in the marginal threshold case.

\section{Special Cases and Consistency Checks}
\label{sec:special}

The purpose is not to rederive the full collapse problem in these theories, but to check that the Lovelock expressions obtained above reduce to the known Einstein, higher-dimensional Einstein, and Einstein--Gauss--Bonnet results for dust collapse~\cite{Eardley:1978tr,Christodoulou:1984mz,Newman:1985gt,Joshi:1993zg,Singh:1994tb,Ghosh:2001fb,Patil:2003yp,Goswami:2002he,Maeda:2006pm,Jhingan:2010zz}.

\subsection{Einstein gravity}
\label{subsec:einstein}

Einstein gravity corresponds to \(N=1\), so the Lovelock polynomial contains only the linear term,
\begin{equation}
P(y)=c_1y .
\end{equation}
In the marginally bound case, Eq.~\eqref{eq:master_X_full} integrates exactly to
\begin{equation}
X(t,r)^{\frac{n+1}{2}}
=
1
-
\frac{n+1}{2}
\left(
\frac{\mu(r)}{c_1}
\right)^{1/2}
t .
\label{eq:Einstein_solution}
\end{equation}
Therefore the singularity curve is
\begin{equation}
t_s(r)
=
\frac{2}{n+1}
\left(
\frac{c_1}{\mu(r)}
\right)^{1/2}.
\label{eq:ts_Einstein}
\end{equation}
This agrees with the general expression~\eqref{eq:ts_exact_sec3} specialized to \(P(y)=c_1y\).

The apparent-horizon equation~\eqref{eq:ah_poly_sec4} reduces to
\begin{equation}
c_1R_{\rm ah}^{\,n-1}
=
M(r).
\label{eq:ah_Einstein}
\end{equation}
Using \(M(r)=r^{n+1}\mu(r)\), one obtains near the center
\begin{equation}
R_{\rm ah}(r)
\sim
\left(
\frac{\mu_0}{c_1}
\right)^{\frac{1}{n-1}}
r^{\frac{n+1}{n-1}} .
\end{equation}
The corresponding trapping-delay exponent is
\begin{equation}
\beta_1
=
\frac{D-1}{D-3}
=
\frac{n+1}{n-1}.
\label{eq:beta_Einstein}
\end{equation}
Thus
\begin{equation}
t_s(r)-t_{\rm ah}(r)
\sim
r^{\beta_1}.
\end{equation}

The general visibility criterion~\eqref{eq:visibility_summary} becomes
\begin{equation}
\ell
<
\frac{D-1}{D-3},
\quad
\chi_\ell>0 .
\label{eq:Einstein_visibility}
\end{equation}
For \(D=4\), one has
\begin{equation}
\beta_1=3 .
\end{equation}
Hence profiles with \(\ell=1,2\) and \(\chi_\ell>0\) give locally visible central singularities, \(\ell=3\) is the threshold case, and \(\ell>3\) is covered. 
This reproduces the standard marginally bound Tolman--Bondi result, where the first and second nonvanishing density derivatives lead to local nakedness, while the third derivative gives a coefficient-dependent threshold~\cite{Joshi:1993zg,Singh:1994tb}.

In higher-dimensional Einstein gravity the criterion becomes more restrictive. 
For \(D=5\), one has \(\beta_1=2\), so smooth quadratic profiles lie at threshold. 
For \(D\geq6\),
\begin{equation}
\beta_1<2 .
\end{equation}
Thus generic smooth profiles with \(\ell=2\) are covered. 
This agrees with the known higher-dimensional Tolman--Bondi result that smooth analytic dust profiles tend to restore black-hole formation in sufficiently high dimensions~\cite{Patil:2003yp,Goswami:2002he}.

The Einstein limit therefore checks three ingredients of the general Lovelock analysis: the exact collapse solution, the apparent-horizon scaling, and the visibility criterion.

\subsection{Einstein--Gauss--Bonnet gravity}
\label{subsec:egb}

Einstein--Gauss--Bonnet gravity corresponds to \(N=2\). 
The Lovelock polynomial is
\begin{equation}
P(y)
=
c_1y+c_2y^2,
\quad
c_2\neq0 .
\label{eq:P_EGB}
\end{equation}
The marginally bound master equation becomes
\begin{equation}
c_1y+c_2y^2
=
\frac{\mu(r)}{X^{n+1}} .
\label{eq:master_EGB}
\end{equation}
On the physical branch with \(c_2>0\), the positive root is
\begin{equation}
y
=
\frac{-c_1+\sqrt{c_1^2+4c_2\mu(r)X^{-n-1}}}{2c_2}.
\label{eq:y_EGB_root}
\end{equation}
Near the singularity \(X\to0\), the quadratic term dominates,
\begin{equation}
P(y)\sim c_2y^2 .
\end{equation}
Therefore
\begin{equation}
\dot X
\sim
-
\left(
\frac{\mu(r)}{c_2}
\right)^{1/4}
X^{-\frac{D-5}{4}} ,
\label{eq:Xdot_EGB_asymp}
\end{equation}
and the near-singularity scaling exponent is
\begin{equation}
\alpha_2
=
\frac{4}{D-1}.
\label{eq:alpha_EGB}
\end{equation}
This is the \(N=2\) specialization of Eq.~\eqref{eq:alphaN}.

The apparent-horizon equation~\eqref{eq:ah_poly_sec4} becomes
\begin{equation}
M(r)
=
c_1R_{\rm ah}^{\,n-1}
+
c_2R_{\rm ah}^{\,n-3}.
\label{eq:ah_EGB}
\end{equation}
The behavior depends on dimension. In \(D=5\), one has \(n=3\), and Eq.~\eqref{eq:ah_EGB} reduces to
\begin{equation}
M(r)
=
c_1R_{\rm ah}^{2}
+
c_2 .
\label{eq:ah_EGB_D5}
\end{equation}
Since \(M(r)\to0\) as \(r\to0\), while \(c_2>0\) on the branch considered here, Eq.~\eqref{eq:ah_EGB_D5} has no solution sufficiently close to the center. 
Thus no apparent horizon forms in a neighborhood of the central shell-focusing singularity. 
This is the critical Lovelock case,
\begin{equation}
D-1-2N=0 .
\end{equation}
The sign assumption is essential. With the opposite sign of the highest Lovelock coefficient, the near-center apparent-horizon equation would have a different behavior; for example, if \(c_2<0\) and \(c_1>0\), then
\begin{equation}
c_1R_{\rm ah}^2+c_2=M(r)
\end{equation}
would give a finite nonzero limiting value for \(R_{\rm ah}\) as \(M(r)\to0\). 

Consequently, if the singularity curve opens outward, \(\chi_\ell>0\), outgoing radial null geodesics can emerge into an untrapped neighborhood. 
This reproduces the characteristic five-dimensional EGB behavior found in dust collapse, where the causal structure differs sharply from the higher-dimensional \(D\geq6\) cases~\cite{Maeda:2006pm}.

For \(D>5\), one has \(D-1-2N=D-5>0\). 
Near the center the Gauss--Bonnet term dominates the apparent-horizon equation, giving
\begin{equation}
R_{\rm ah}(r)
\sim
\left(
\frac{M(r)}{c_2}
\right)^{\frac{1}{D-5}} .
\end{equation}
Using \(M(r)=r^{D-1}\mu(r)\), this becomes
\begin{equation}
R_{\rm ah}(r)
\sim
\left(
\frac{\mu_0}{c_2}
\right)^{\frac{1}{D-5}}
r^{\frac{D-1}{D-5}} .
\label{eq:Rah_EGB}
\end{equation}
The trapping-delay exponent is therefore
\begin{equation}
\beta_2
=
\frac{D-1}{D-5}.
\label{eq:beta_EGB}
\end{equation}
Since
\begin{equation}
\beta_2
>
\beta_1
=
\frac{D-1}{D-3},
\end{equation}
the Gauss--Bonnet term delays the approach of the apparent horizon to the central singularity relative to Einstein gravity in the same dimension.

The visibility criterion becomes
\begin{equation}
\ell
<
\frac{D-1}{D-5},
\quad
\chi_\ell>0 .
\label{eq:EGB_visibility}
\end{equation}
For smooth generic data, \(\ell=2\), so the noncritical EGB visibility condition is satisfied only for \(D=6,7,8\), while \(D=5\) is the critical case and \(D\geq9\) is locally covered. 
This dimensional window is useful when comparing with pure Gauss--Bonnet collapse, where recent analyses find covered smooth-profile collapse outside the locally visible low-dimensional window~\cite{Dialektopoulos:2023qda,Kumar:2025qqj}.

Thus the EGB term enlarges the range of initial density profiles leading to local visibility. For example, in \(D=6\),
\begin{equation}
\beta_2=5,
\end{equation}
so smooth profiles with \(\ell=2\) satisfy the visibility condition. 
This contrasts with six-dimensional Einstein gravity, for which \(\beta_1=5/3\) and smooth quadratic profiles are covered.

This is consistent with the known EGB dust-collapse analysis, where the five-dimensional case and the \(D\geq6\) cases have qualitatively different apparent-horizon and causal structures~\cite{Maeda:2006pm}.

The EGB limit therefore illustrates the main mechanism found in the general Lovelock analysis. 
The highest Lovelock term controls the near-singularity dynamics, delays the formation of trapped surfaces, and, in the critical odd-dimensional case, prevents the apparent horizon from forming near the center.

\section{Extension to Bound and Unbound Collapse}
\label{sec:nonmarginal}

We now discuss how the marginally bound analysis is modified when the energy function is nonzero. 
With the convention of Eq.~\eqref{eq:ltb_metric}, the non-marginal cases are described by
\begin{equation}
E(r)=r^2b(r),
\end{equation}
with \(b(r)<0\) corresponding to bound collapse and \(b(r)>0\) to unbound collapse. 
The master equation~\eqref{eq:master_X_full} can be written as
\begin{equation}
P(y)
=
\frac{\mu(r)}{X^{n+1}},
\quad
y
\equiv
\frac{\dot X^2-b(r)}{X^2}.
\label{eq:nonmarginal_y}
\end{equation}
Therefore
\begin{equation}
\dot X^2
=
X^2y+b(r).
\label{eq:Xdot_nonmarginal}
\end{equation}
On the collapsing branch,
\begin{equation}
\dot X
=
-\sqrt{X^2y+b(r)}.
\label{eq:Xdot_nonmarginal_branch}
\end{equation}
Thus the sign of \(b(r)\) has the expected effect: positive \(b(r)\) increases the collapse velocity, while negative \(b(r)\) decreases it.

\subsection{Generic singularity time for \(b(r)\neq0\)}
\label{subsec:nonmarginal_ts}

Let \(Y(X;\mu)\) denote the positive root of
\begin{equation}
P(Y)
=
\frac{\mu}{X^{n+1}} .
\label{eq:Y_def_nonmarginal}
\end{equation}
Then, for a shell labeled by \(r\),
\begin{equation}
y(X,r)=Y(X;\mu(r)).
\end{equation}
The singularity time is therefore
\begin{equation}
t_s(r)
=
\int_0^1
\frac{dX}{\sqrt{X^2Y(X;\mu(r))+b(r)}} .
\label{eq:ts_nonmarginal}
\end{equation}
This expression replaces the marginally bound formula~\eqref{eq:ts_exact_sec3}. 
In contrast with the case \(b=0\), the integral cannot be reduced to a purely \(y\)-dependent expression, because the term \(b(r)\) remains explicitly in the denominator.

Near the center we write
\begin{equation}
\mu(r)
=
\mu_0+\mu_\ell r^\ell+O(r^{\ell+1}),
\quad
\ell\geq1,
\label{eq:mu_expand_nonmarginal}
\end{equation}
and
\begin{equation}
b(r)
=
b_0+b_p r^p+O(r^{p+1}),
\quad
p\geq1.
\label{eq:b_expand_nonmarginal}
\end{equation}
The central singularity time is
\begin{equation}
t_0
=
\int_0^1
\frac{dX}{\sqrt{X^2Y(X;\mu_0)+b_0}} ,
\label{eq:t0_nonmarginal}
\end{equation}
assuming that the denominator is positive along the collapsing branch.

It is useful to regard the singularity time as a function
\begin{equation}
\mathcal{T}(\mu,b)
=
\int_0^1
\frac{dX}{\sqrt{X^2Y(X;\mu)+b}} .
\label{eq:T_mu_b}
\end{equation}
Then
\begin{equation}
t_s(r)
=
\mathcal{T}(\mu(r),b(r)).
\end{equation}
Expanding near \(r=0\), one obtains
\begin{equation}
t_s(r)
=
t_0
+
\mathcal{T}_{,\mu}\big|_0\,\mu_\ell r^\ell
+
\mathcal{T}_{,b}\big|_0\,b_p r^p
+
\cdots ,
\label{eq:ts_nonmarginal_expand}
\end{equation}
where the derivatives are evaluated at \((\mu,b)=(\mu_0,b_0)\).

The derivative with respect to \(b\) is
\begin{equation}
\mathcal{T}_{,b}\big|_0
=
-\frac{1}{2}
\int_0^1
\frac{dX}
{\left[X^2Y(X;\mu_0)+b_0\right]^{3/2}}
<0 .
\label{eq:T_b}
\end{equation}
Thus increasing \(b\) decreases the collapse time, while decreasing \(b\) increases it. 
The derivative with respect to \(\mu\) is also negative on the physical branch \(P'(Y)>0\). Indeed, differentiating Eq.~\eqref{eq:Y_def_nonmarginal} gives
\begin{equation}
\frac{\partial Y}{\partial \mu}
=
\frac{1}{X^{n+1}P'(Y)},
\end{equation}
and hence
\begin{equation}
\mathcal{T}_{,\mu}\big|_0
=
-\frac{1}{2}
\int_0^1
\frac{
X^{1-n}\,dX
}{
P'(Y)
\left[
X^2Y(X;\mu_0)+b_0
\right]^{3/2}
}
<0 .
\label{eq:T_mu}
\end{equation}
Therefore a lower density away from the center, \(\mu_\ell<0\), tends to delay the singularity formation of neighboring shells, as in the marginally bound case.

Both derivatives are finite under the assumptions above. In particular, near \(X=0\),
\begin{equation}
Y(X;\mu_0)
\sim
\left(
\frac{\mu_0}{c_N}
\right)^{1/N}
X^{-\frac{D-1}{N}},
\end{equation}
and the integrand in Eq.~\eqref{eq:T_mu} behaves as
\begin{equation}
X^{-1+\frac{D-1}{2N}},
\end{equation}
which is integrable for the Lovelock orders considered here. Thus the expansion of \(t_s(r)\) in powers of the initial data is well defined.

The leading opening of the singularity curve is determined by the first nonzero contribution in Eq.~\eqref{eq:ts_nonmarginal_expand}. We write
\begin{equation}
t_s(r)
=
t_0+\chi_\eta r^\eta+O(r^{\eta+1}),
\quad
\chi_\eta\neq0,
\label{eq:ts_eta_nonmarginal}
\end{equation}
where \(\eta\) is the smallest power with a nonzero coefficient after the density and energy contributions are combined.

This gives a useful distinction between density-driven and energy-driven openings of the singularity curve. If \(\ell<p\), then
\begin{equation}
\eta=\ell,
\quad
\chi_\eta
=
\mathcal{T}_{,\mu}\big|_0\,\mu_\ell .
\end{equation}
If \(p<\ell\), then
\begin{equation}
\eta=p,
\quad
\chi_\eta
=
\mathcal{T}_{,b}\big|_0\,b_p .
\end{equation}
Since \(\mathcal{T}_{,b}\big|_0<0\), an energy-dominated opening satisfies
\begin{equation}
\chi_\eta>0
\quad
\Longleftrightarrow
\quad
b_p<0 .
\end{equation}
Thus an inhomogeneous energy function can independently delay the collapse of neighboring shells and source local visibility, even when the density profile is homogeneous to the same order.

If \(\ell=p\), then
\begin{equation}
\chi_\eta
=
\mathcal{T}_{,\mu}\big|_0\,\mu_\ell
+
\mathcal{T}_{,b}\big|_0\,b_p .
\end{equation}
A cancellation between the density and energy contributions can make this coefficient vanish, in which case the leading power \(\eta\) is pushed to higher order.

For smooth data at the center, both the density profile and the energy function typically begin at quadratic order:
\begin{equation}
\ell=2,
\quad
p=2.
\end{equation}
The positivity of the density is unchanged by the energy function \(b(r)\), since the density is determined by the mass function \(M(r)\) through Eq.~\eqref{eq:rho_general}. Thus the near-center positivity condition discussed in Sec.~\ref{subsec:regularity} carries over directly to the non-marginal case.

\subsection{Near-singularity behavior for \(D-1-2N>0\)}
\label{subsec:nonmarginal_noncritical}

We now consider the noncritical branch
\begin{equation}
s
\equiv
D-1-2N
>
0 .
\end{equation}
Near the shell-focusing singularity, \(X\to0\), the Lovelock polynomial is dominated by the highest nonvanishing order:
\begin{equation}
P(y)
\sim
c_Ny^N .
\end{equation}
Using Eq.~\eqref{eq:nonmarginal_y}, this gives
\begin{equation}
y
\sim
\left(
\frac{\mu(r)}{c_N}
\right)^{1/N}
X^{-\frac{n+1}{N}} .
\label{eq:y_nonmarginal_asymp}
\end{equation}
Therefore
\begin{equation}
X^2y
\sim
\left(
\frac{\mu(r)}{c_N}
\right)^{1/N}
X^{-\frac{D-1-2N}{N}} .
\label{eq:X2y_noncritical}
\end{equation}
Since \(D-1-2N>0\), the quantity \(X^2y\) diverges as \(X\to0\). 
Therefore a finite \(b(r)\) cannot obstruct the final approach to \(X=0\). Indeed, if we define
\begin{equation}
f(X)\equiv X^2Y(X;\mu).
\end{equation}
Differentiating Eq.~\eqref{eq:Y_def_nonmarginal} gives
\begin{equation}
P'(Y)\frac{dY}{dX}
=
-\frac{n+1}{X}P(Y),
\end{equation}
and hence
\begin{equation}
f'(X)
=
X
\left[
2Y
-
(n+1)\frac{P(Y)}{P'(Y)}
\right].
\label{eq:fprime_nonmarginal}
\end{equation}
Thus \(f'(X)<0\) whenever
\begin{equation}
\frac{2YP'(Y)}{P(Y)}
<
n+1 .
\label{eq:monotonic_condition}
\end{equation}
If the Lovelock coefficients on the physical branch satisfy \(c_m\geq0\), then
\begin{equation}
\frac{YP'(Y)}{P(Y)}
=
\frac{\sum_{m=1}^{N}m c_mY^m}{\sum_{m=1}^{N}c_mY^m}
\leq N .
\end{equation}
Since \(D-1-2N>0\), equivalently \(2N<n+1\), Eq.~\eqref{eq:monotonic_condition} holds. Therefore \(f'(X)<0\) on the interval of collapse: \(X^2Y\) increases monotonically as \(X\) decreases toward zero. In that case,
\begin{equation}
\dot X^2=X^2Y+b(r)
\end{equation}
is minimized at the initial surface \(X=1\). Hence an initially collapsing shell with \(\dot X(0,r)^2>0\) cannot encounter a turning point before reaching the shell-focusing singularity.

If the Lovelock coefficients are not all nonnegative, this global monotonicity need not hold. However, the near-singularity statement remains valid under the weaker assumption \(c_N>0\). Indeed,
\begin{equation}
\frac{YP'(Y)}{P(Y)}
\longrightarrow
N
\quad
(Y\to\infty),
\end{equation}
and therefore Eq.~\eqref{eq:monotonic_condition} holds sufficiently close to \(X=0\). Thus, independently of the lower-order Lovelock coefficients, a finite \(b(r)\) cannot obstruct the final approach to the singularity in the noncritical branch.

Thus the finite term \(b(r)\) in Eq.~\eqref{eq:Xdot_nonmarginal} is subleading:
\begin{equation}
X^2y+b(r)
\sim
X^2y .
\end{equation}
The leading collapse dynamics is therefore the same as in the marginally bound case:
\begin{equation}
\dot X
\sim
-
\left(
\frac{\mu(r)}{c_N}
\right)^{\frac{1}{2N}}
X^{-\frac{D-1-2N}{2N}} .
\label{eq:Xdot_nonmarginal_noncritical}
\end{equation}
Integrating, one obtains
\begin{equation}
X(t,r)
\sim
A(r)
\left[
t_s(r)-t
\right]^{\alpha_N},
\quad
\alpha_N
=
\frac{2N}{D-1},
\label{eq:X_scaling_nonmarginal}
\end{equation}
with the same exponent as in the marginally bound case.

The apparent-horizon condition is also unchanged. 
Indeed, the condition \(g^{ab}\partial_aR\partial_bR=0\) gives
\begin{equation}
\dot R^2=1+E(r),
\end{equation}
or equivalently
\begin{equation}
\dot R^2-E(r)=1.
\end{equation}
Substituting this into the Lovelock--LTB master equation gives again
\begin{equation}
\sum_{m=1}^{N}
c_m
R_{\rm ah}^{\,n+1-2m}
=
M(r).
\label{eq:ah_nonmarginal}
\end{equation}
Therefore, for \(D-1-2N>0\),
\begin{equation}
R_{\rm ah}(r)
\sim
\left(
\frac{\mu_0}{c_N}
\right)^{1/(D-1-2N)}
r^{\beta_N},
\quad
\beta_N
=
\frac{D-1}{D-1-2N}.
\label{eq:Rah_nonmarginal}
\end{equation}
The time difference between singularity formation and apparent-horizon formation is
\begin{equation}
t_s(r)-t_{\rm ah}(r)
=
\int_0^{X_{\rm ah}(r)}
\frac{dX}
{\sqrt{X^2Y(X;\mu(r))+b(r)}} .
\label{eq:time_diff_nonmarginal}
\end{equation}
Since \(b(r)\) is subleading near \(X=0\), this gives the same leading scaling:
\begin{equation}
t_s(r)-t_{\rm ah}(r)
\sim
r^{\beta_N}.
\label{eq:time_diff_nonmarginal_scaling}
\end{equation}

The local-visibility analysis is therefore unchanged in form, but the relevant power is not necessarily \(\ell\). 
It is the leading power \(\eta\) in the full singularity curve~\eqref{eq:ts_eta_nonmarginal}. 
Thus, for \(D-1-2N>0\), the leading criterion becomes
\begin{equation}
\begin{cases}
\text{locally visible},
&
\eta<\beta_N
\quad\text{and}\quad
\chi_\eta>0,
\\[5pt]
\text{threshold case},
&
\eta=\beta_N,
\\[5pt]
\text{covered},
&
\eta>\beta_N .
\end{cases}
\label{eq:visibility_nonmarginal}
\end{equation}
Since \(\eta\) is an integer for smooth expansions of the initial data, the threshold case \(\eta=\beta_N\) can occur only when \(\beta_N\) is a positive integer. Otherwise the leading-order classification reduces effectively to the two cases \(\eta<\beta_N\) and \(\eta>\beta_N\).
In the marginally bound case, \(\eta=\ell\), and Eq.~\eqref{eq:visibility_nonmarginal} reduces to the criterion derived in Sec.~\ref{subsec:criteria}. 
For nonzero \(b(r)\), the energy profile can modify the opening of the singularity curve through Eq.~\eqref{eq:ts_nonmarginal_expand}; however, it does not change the universal exponents \(\alpha_N\) and \(\beta_N\) in the noncritical branch.

The curvature-strength analysis also carries over directly. In the noncritical branch, \(b(r)\) is subleading in the near-singularity velocity, and the local form of the areal radius is the same as in the marginally bound case, with \(\ell\) replaced by \(\eta\):
\begin{equation}
R(t,r)
\sim
rA_0\,[t_s(r)-t]^{\alpha_N}.
\end{equation}
Along a locally outgoing null ray one has
\begin{equation}
t_s(r)-t(r)
\sim
\chi_\eta r^\eta,
\end{equation}
and therefore the affine parameter and the Ricci contraction scale as
\begin{equation}
\lambda
\sim
r^{1+\alpha_N\eta},
\quad
R_{\mu\nu}k^\mu k^\nu
\sim
r^{-2\eta}.
\end{equation}
Equivalently,
\begin{equation}
R_{\mu\nu}k^\mu k^\nu
\sim
\lambda^{-q},
\quad
q
=
\frac{2\eta}{1+\alpha_N\eta}.
\label{eq:q_nonmarginal_noncritical}
\end{equation}
For the strict visible branch \(\eta<\beta_N\), this gives \(1<q<2\). Thus the non-marginal locally visible singularities are Królak-strong along the outgoing null rays. At the threshold \(\eta=\beta_N\), one obtains \(q=2\), so a visible threshold ray satisfies the corresponding Tipler Ricci-focusing condition.

Thus, in the noncritical branch, the energy function does not change the universal Lovelock exponents \(\alpha_N\) and \(\beta_N\). Its role is to modify the opening of the singularity curve, replacing \(\ell\) by the effective leading power \(\eta\). Consequently, the curvature-strength exponent \eqref{eq:q_nonmarginal_noncritical} can change through the energy-dependent value of \(\eta\).

\subsection{Critical case \(D-1-2N=0\)}
\label{subsec:nonmarginal_critical}

The critical case requires separate treatment. 
When
\begin{equation}
D-1-2N=0,
\end{equation}
the asymptotic behavior~\eqref{eq:y_nonmarginal_asymp} gives
\begin{equation}
X^2y
\sim
\left(
\frac{\mu(r)}{c_N}
\right)^{1/N}.
\end{equation}
Thus \(X^2y\) no longer diverges near \(X=0\). 
The energy function \(b(r)\) contributes at the same order as the Lovelock term in the collapse velocity:
\begin{equation}
\dot X^2
\sim
\left(
\frac{\mu(r)}{c_N}
\right)^{1/N}
+
b(r).
\label{eq:Xdot_critical_b}
\end{equation}
On the positive-coupling branch considered above, the function $f(X;\mu)=X^2Y(X;\mu)$ is nonincreasing as a function of \(X\), and therefore nondecreasing along the collapse as \(X\) decreases. In the critical case it approaches the finite limit
\begin{align}
f(0;\mu)
=
\left(\frac{\mu}{c_N}\right)^{1/N}.
\end{align}
Moreover,
\begin{align}
f(1;\mu)=Y(1;\mu)
\le
\left(\frac{\mu}{c_N}\right)^{1/N},
\end{align}
with equality in pure Lovelock gravity. Hence an initially collapsing shell
with
\begin{align}
\dot X^2(1,r)=f(1;\mu(r))+b(r)>0
\end{align}
cannot encounter a turning point before reaching \(X=0\). If
\begin{align}
b_0<-\left(\frac{\mu_0}{c_N}\right)^{1/N},
\end{align}
then the terminal velocity squared is negative; on the positive-coupling
branch this is not a collapsing solution that bounces, but data outside the
branch that reaches the shell-focusing singularity. A genuine bounce of
initially collapsing data would require a nonmonotonic \(f(X;\mu)\), which
can occur only for more general Lovelock couplings or branches not assumed
here.

The central admissibility condition is
\begin{align}
V_0^2
=
\left(\frac{\mu_0}{c_N}\right)^{1/N}+b_0
>0 .
\end{align}
More generally, for each shell one requires
\begin{align}
V^2(r)
\equiv
\left(\frac{\mu(r)}{c_N}\right)^{1/N}+b(r)>0 .
\end{align}
Consequently, the collapse velocity near the singularity is
\begin{equation}
\dot X
\sim
-
V(r),
\quad
V(r)
\equiv
\left[
\left(
\frac{\mu(r)}{c_N}
\right)^{1/N}
+
b(r)
\right]^{1/2}.
\label{eq:V_critical}
\end{equation}
Therefore
\begin{equation}
X(t,r)
\sim
V(r)\,[t_s(r)-t].
\label{eq:X_critical_nonmarginal}
\end{equation}
The exponent remains
\begin{equation}
\alpha_N=1,
\end{equation}
as in the marginally bound critical case, but the coefficient now depends on the energy profile \(b(r)\).

The apparent-horizon equation remains Eq.~\eqref{eq:ah_nonmarginal}. 
For \(D-1-2N=0\), its leading term near the center is
\begin{equation}
c_N+\cdots=M(r).
\end{equation}
Since \(M(r)\to0\) as \(r\to0\), while \(c_N\neq0\), there is no apparent-horizon solution sufficiently close to the center. 
Thus the absence of trapped surfaces near the central shell-focusing singularity is unchanged by the nonzero energy function.

The outgoing radial null analysis also keeps the same structure, with the singularity curve written as
\begin{equation}
t_s(r)
=
t_0+\chi_\eta r^\eta+O(r^{\eta+1}).
\end{equation}
Since \(\alpha_N=1\), the null-ray matching gives
\begin{equation}
\gamma
=
1+\eta
>
\eta .
\end{equation}
Therefore, if
\begin{equation}
\chi_\eta>0,
\end{equation}
future-directed outgoing radial null geodesics can emerge from the central singularity into an untrapped neighborhood. 
In the critical Lovelock case, nonzero \(b(r)\) changes the coefficients in the singularity curve and in the near-singularity velocity, but it does not restore local trapped surfaces near the center.

The curvature-strength scaling is also unchanged at the level of exponents. Along a locally outgoing null ray,
\begin{equation}
t_s(r)-t(r)
\sim
\chi_\eta r^\eta,
\end{equation}
and since \(R\sim rV_0[t_s(r)-t]\), one obtains
\begin{equation}
\lambda
\sim
r^{1+\eta},
\quad
R_{\mu\nu}k^\mu k^\nu
\sim
r^{-2\eta}.
\end{equation}
Thus
\begin{equation}
R_{\mu\nu}k^\mu k^\nu
\sim
\lambda^{-q},
\quad
q
=
\frac{2\eta}{1+\eta}.
\label{eq:q_nonmarginal_critical}
\end{equation}
For every finite \(\eta\geq1\),
\begin{equation}
1\leq q<2 .
\end{equation}
Hence, whenever the critical non-marginal collapse forms a locally visible central singularity, that singularity is Królak-strong along the outgoing radial null ray, while the Tipler Ricci-focusing condition is not reached for finite \(\eta\).

We conclude that non-marginal collapse preserves the universal Lovelock exponents \(\alpha_N\) and \(\beta_N\) in the noncritical branch. There, the energy function affects the visibility criterion only through the opening of the singularity curve, replacing \(\ell\) by the effective power \(\eta\). In the critical branch, \(b(r)\) enters the leading terminal velocity itself, while the collapse exponent remains \(\alpha_N=1\). On the standard positive-coupling branch, initially collapsing data that are real at \(X=1\) remain real down to \(X=0\); the energy function changes the terminal velocity but does not generate a bounce before the central
shell-focusing singularity. When a locally visible singularity does form, its curvature-strength scaling is the same as in the marginally bound case, with \(\ell\) replaced by \(\eta\).

\section{Conclusions}
\label{sec:conclusions}

We have studied spherical dust collapse in general Lovelock gravity in arbitrary spacetime dimension, focusing on the local visibility of central shell-focusing singularities. 
The result is counterintuitive from the perspective of cosmic censorship: on the collapse branch considered here, higher-curvature Lovelock terms do not hide the central singularity more efficiently. 
Instead, the highest Lovelock term delays the formation of trapped surfaces, and in the critical odd-dimensional case removes them from a sufficiently small neighborhood of the center.

The central mechanism is the universality of the near-singularity dynamics. 
If \(N\) is the highest nonvanishing Lovelock order, then the approach to the shell-focusing singularity is governed only by that order, with
\begin{equation}
\alpha_N
=
\frac{2N}{D-1}.
\end{equation}
Lower-order Lovelock terms affect coefficients and threshold cases, but not the leading exponent. 
Consequently, away from threshold, pure Lovelock collapse and Einstein--Lovelock collapse with the same highest order \(N\) have the same leading local visibility criterion.

The apparent-horizon analysis gives the corresponding trapping-delay exponent
\begin{equation}
\beta_N
=
\frac{D-1}{D-1-2N},
\quad
D-1-2N>0 .
\end{equation}
This exponent increases with \(N\) at fixed dimension. 
Thus higher Lovelock order delays the onset of trapping near the center. 
For a singularity curve
\begin{equation}
t_s(r)
=
t_0+\chi_\ell r^\ell+O(r^{\ell+1}),
\end{equation}
the leading noncritical visibility criterion is
\begin{equation}
\ell<\beta_N,
\quad
\chi_\ell>0 .
\end{equation}
The same exponent \(\beta_N=1/(1-\alpha_N)\) appears independently in the outgoing radial null-geodesic matching, showing that the trapping calculation and the null-ray calculation are controlled by the same local scaling.

This also clarifies the dimensional pattern suggested by the third-order analysis of \cite{Zhou:2011vz}. Their special $D=7$ case is precisely the $N=3$ instance of the critical Lovelock dimension $D=2N+1$. Our result shows that the same mechanism holds for arbitrary highest Lovelock order: the critical odd-dimensional case removes local trapping near the center, whereas for $D>2N+1$ trapping returns with a delay controlled by $\beta_N$.

The locally visible singularities found here are curvature singularities in the usual strong-focusing sense. 
Along outgoing radial null rays, the Ricci contraction diverges with a power fixed by the same near-center data. 
In the strict visible branch this gives Królak-strong behavior, while the Tipler Ricci-focusing condition is reached at the threshold case. 
Thus the local visibility is not a coordinate artifact or a consequence of a curvature-regular limiting geometry.

The extension to non-marginal collapse preserves the same picture in the noncritical branch. 
The energy function \(E(r)=r^2b(r)\) changes the opening of the singularity curve and hence replaces \(\ell\) by the first nonvanishing power \(\eta\) in the full expansion of \(t_s(r)\), but it does not change the universal Lovelock exponents. 
In the critical branch, \(b(r)\) enters the leading collapse velocity itself. 
On the standard positive-coupling branch, however, initially collapsing data that are real at the initial surface remain real down to \(X=0\); sufficiently bound choices that make the terminal velocity imaginary are outside the initially collapsing branch reaching the center, rather than collapsing solutions that bounce before the singularity.

Our analysis is local. 
It establishes the existence of outgoing null rays from the central singularity in an untrapped neighborhood, but it does not decide whether those rays reach the surface of the cloud or future null infinity after matching to an exterior Lovelock spacetime. 
That global question depends on the finite-radius matter profile, on the exterior branch, and on the matching conditions. 
The local result is nevertheless clear: on the branch studied here, higher Lovelock curvature terms enlarge, rather than shrink, the class of dust-collapse data producing locally visible central singularities.

\section*{Acknowledgments}

The work of R.G. is supported by ANID FONDECYT Regular No. 1262002 (Chile). A.K. thanks IUCAA for the hospitality, where part of the work was done. We would also like to dedicate this work to the memory of our friend and colleague Naresh Dadhich, whose deep contributions to Lovelock gravity and higher-curvature theories have strongly shaped this field. His scientific insight and generosity will be warmly remembered. 

\bibliographystyle{apsrev4-2}
\bibliography{references}

@article{Lovelock:1971yv,
    author = "Lovelock, D.",
    title = "{The Einstein tensor and its generalizations}",
    doi = "10.1063/1.1665613",
    journal = "J. Math. Phys.",
    volume = "12",
    pages = "498--501",
    year = "1971"
}

@article{Zumino:1985dp,
    author = "Zumino, Bruno",
    title = "{Gravity Theories in More Than Four-Dimensions}",
    reportNumber = "UCB-PTH-85-13, LBL-19302",
    doi = "10.1016/0370-1573(86)90076-1",
    journal = "Phys. Rept.",
    volume = "137",
    pages = "109",
    year = "1986"
}

@article{Charmousis:2008kc,
    author = "Charmousis, Christos",
    editor = "Papantonopoulos, Eleftherios",
    title = "{Higher order gravity theories and their black hole solutions}",
    eprint = "0805.0568",
    archivePrefix = "arXiv",
    primaryClass = "gr-qc",
    reportNumber = "LPT-08-42",
    doi = "10.1007/978-3-540-88460-6_8",
    journal = "Lect. Notes Phys.",
    volume = "769",
    pages = "299--346",
    year = "2009"
}

@book{Joshi:2008zz,
    editor = "Joshi, Pankaj S.",
    title = "{Gravitational Collapse and Spacetime Singularities}",
    doi = "10.1017/CBO9780511536274",
    isbn = "978-1-107-40536-3, 978-0-521-87104-4, 978-0-511-37283-4",
    publisher = "Cambridge University Press",
    series = "Cambridge Monographs on Mathematical Physics",
    month = "9",
    year = "2012"
}

@article{Hayward:1993wb,
    author = "Hayward, S. A.",
    title = "{General laws of black hole dynamics}",
    doi = "10.1103/PhysRevD.49.6467",
    journal = "Phys. Rev. D",
    volume = "49",
    pages = "6467--6474",
    year = "1994"
}

@article{Booth:2005qc,
    author = "Booth, Ivan",
    title = "{Black hole boundaries}",
    eprint = "gr-qc/0508107",
    archivePrefix = "arXiv",
    doi = "10.1139/p05-063",
    journal = "Can. J. Phys.",
    volume = "83",
    pages = "1073--1099",
    year = "2005"
}

@article{Maeda:2006pm,
    author = "Maeda, Hideki",
    title = "{Final fate of spherically symmetric gravitational collapse of a dust cloud in Einstein-Gauss-Bonnet gravity}",
    eprint = "gr-qc/0602109",
    archivePrefix = "arXiv",
    doi = "10.1103/PhysRevD.73.104004",
    journal = "Phys. Rev. D",
    volume = "73",
    pages = "104004",
    year = "2006"
}

@article{Ohashi:2011zza,
    author = "Ohashi, Seiju and Shiromizu, Tetsuya and Jhingan, Sanjay",
    title = "{Spherical collapse of inhomogeneous dust cloud in the Lovelock theory}",
    eprint = "1103.3826",
    archivePrefix = "arXiv",
    primaryClass = "gr-qc",
    doi = "10.1103/PhysRevD.84.024021",
    journal = "Phys. Rev. D",
    volume = "84",
    pages = "024021",
    year = "2011"
}

@article{Jhingan:2010zz,
    author = "Jhingan, S. and Ghosh, Sushant G.",
    title = "{Inhomogeneous dust collapse in D-5 Einstein-Gauss-Bonnet gravity}",
    eprint = "1002.3245",
    archivePrefix = "arXiv",
    primaryClass = "gr-qc",
    doi = "10.1103/PhysRevD.81.024010",
    journal = "Phys. Rev. D",
    volume = "81",
    pages = "024010",
    year = "2010"
}

@article{Tipler:1977zza,
    author = "Tipler, Frank J.",
    title = "{Singularities in conformally flat spacetimes}",
    doi = "10.1016/0375-9601(77)90508-4",
    journal = "Phys. Lett. A",
    volume = "64",
    pages = "8--10",
    year = "1977"
}

@article{Clarke:1985,
    author = "Clarke, C. J. S. and Krolak, A.",
    title = "{Conditions for the occurrence of strong curvature singularities}",
    journal = "J. Geom. Phys.",
    volume = "2",
    pages = "127--143",
    year = "1985",
    doi = "10.1016/0393-0440(85)90012-9"
}

@article{Joshi:1996qc,
    author = "Joshi, Pankaj S. and Krolak, Andrzej",
    title = "{Naked strong curvature singularities in Szekeres space-times}",
    eprint = "gr-qc/9605033",
    archivePrefix = "arXiv",
    doi = "10.1088/0264-9381/13/11/020",
    journal = "Class. Quant. Grav.",
    volume = "13",
    pages = "3069--3074",
    year = "1996"
}

@article{Joshi:1993zg,
    author = "Joshi, P. S. and Dwivedi, I. H.",
    title = "{Naked singularities in spherically symmetric inhomogeneous Tolman-Bondi dust cloud collapse}",
    eprint = "gr-qc/9303037",
    archivePrefix = "arXiv",
    reportNumber = "TIFR-TAP-9-92",
    doi = "10.1103/PhysRevD.47.5357",
    journal = "Phys. Rev. D",
    volume = "47",
    pages = "5357--5369",
    year = "1993"
}

@article{Singh:1994tb,
    author = "Singh, T. P. and Joshi, P. S.",
    title = "{The Final fate of spherical inhomogeneous dust collapse}",
    eprint = "gr-qc/9409062",
    archivePrefix = "arXiv",
    doi = "10.1088/0264-9381/13/3/019",
    journal = "Class. Quant. Grav.",
    volume = "13",
    pages = "559--572",
    year = "1996"
}

@article{Patil:2003yp,
    author = "Patil, K. D.",
    title = "{Structure of radial null geodesics in higher dimensional dust collapse}",
    doi = "10.1103/PhysRevD.67.024017",
    journal = "Phys. Rev. D",
    volume = "67",
    pages = "024017",
    year = "2003"
}

@article{Goswami:2002he,
    author = "Goswami, Rituparno and Joshi, Pankaj S.",
    title = "{Spherical dust collapse in higher dimensions}",
    eprint = "gr-qc/0212097",
    archivePrefix = "arXiv",
    doi = "10.1103/PhysRevD.69.044002",
    journal = "Phys. Rev. D",
    volume = "69",
    pages = "044002",
    year = "2004"
}

@article{Oppenheimer:1939ue,
    author = "Oppenheimer, J. R. and Snyder, H.",
    title = "{On Continued gravitational contraction}",
    doi = "10.1103/PhysRev.56.455",
    journal = "Phys. Rev.",
    volume = "56",
    pages = "455--459",
    year = "1939"
}

@article{Penrose:1969pc,
    author = "Penrose, R.",
    title = "{Gravitational collapse: The role of general relativity}",
    doi = "10.1023/A:1016578408204",
    journal = "Riv. Nuovo Cim.",
    volume = "1",
    pages = "252--276",
    year = "1969"
}

@article{Eardley:1978tr,
    author = "Eardley, Douglas M. and Smarr, Larry",
    title = "{Time function in numerical relativity. Marginally bound dust collapse}",
    doi = "10.1103/PhysRevD.19.2239",
    journal = "Phys. Rev. D",
    volume = "19",
    pages = "2239--2259",
    year = "1979"
}

@article{Christodoulou:1984mz,
    author = "Christodoulou, Demetrios",
    title = "{Violation of cosmic censorship in the gravitational collapse of a dust cloud}",
    doi = "10.1007/BF01223743",
    journal = "Commun. Math. Phys.",
    volume = "93",
    pages = "171--195",
    year = "1984"
}

@article{Newman:1985gt,
    author = "Newman, Richard P. A. C.",
    title = "{Strengths of naked singularities in Tolman-Bondi space-times}",
    doi = "10.1088/0264-9381/3/4/007",
    journal = "Class. Quant. Grav.",
    volume = "3",
    pages = "527--539",
    year = "1986"
}

@article{Ghosh:2001fb,
    author = "Ghosh, S. G. and Beesham, A.",
    title = "{Higher dimensional inhomogeneous dust collapse and cosmic censorship}",
    eprint = "gr-qc/0108011",
    archivePrefix = "arXiv",
    doi = "10.1103/PhysRevD.64.124005",
    journal = "Phys. Rev. D",
    volume = "64",
    pages = "124005",
    year = "2001"
}

@article{Zhou:2011vz,
    author = "Zhou, Kang and Yang, Zhan-Ying and Zou, De-Cheng and Yue, Rui-Hong",
    title = "{Spherically symmetric gravitational collapse of a dust cloud in third order Lovelock Gravity}",
    eprint = "1107.2730",
    archivePrefix = "arXiv",
    primaryClass = "gr-qc",
    doi = "10.1142/S0218271811020408",
    journal = "Int. J. Mod. Phys. D",
    volume = "20",
    pages = "2317--2335",
    year = "2011"
}

@article{Ohashi:2012wfa,
    author = "Ohashi, Seiju and Shiromizu, Tetsuya and Jhingan, Sanjay",
    title = "{Gravitational collapse of charged dust cloud in the Lovelock gravity}",
    eprint = "1205.5363",
    archivePrefix = "arXiv",
    primaryClass = "gr-qc",
    doi = "10.1103/PhysRevD.86.044008",
    journal = "Phys. Rev. D",
    volume = "86",
    pages = "044008",
    year = "2012"
}

@article{Dadhich:2013bya,
    author = "Dadhich, Naresh and Ghosh, Sushant G. and Jhingan, Sanjay",
    title = "{Gravitational collapse in pure Lovelock gravity in higher dimensions}",
    eprint = "1308.4312",
    archivePrefix = "arXiv",
    primaryClass = "gr-qc",
    doi = "10.1103/PhysRevD.88.084024",
    journal = "Phys. Rev. D",
    volume = "88",
    pages = "084024",
    year = "2013"
}

@article{Dialektopoulos:2023qda,
    author = "Dialektopoulos, Konstantinos F. and Malafarina, Daniele and Dadhich, Naresh",
    title = "{Gravitational collapse in pure Gauss-Bonnet gravity}",
    eprint = "2306.10872",
    archivePrefix = "arXiv",
    primaryClass = "gr-qc",
    doi = "10.1103/PhysRevD.108.044080",
    journal = "Phys. Rev. D",
    volume = "108",
    number = "4",
    pages = "044080",
    year = "2023"
}

@article{Kumar:2025qqj,
    author = "Kumar, Akshay and Chatterjee, Ayan and Jaryal, Suresh C.",
    title = "{Gravitational collapse in pure Gauss{\textendash}Bonnet theory}",
    eprint = "2505.05205",
    archivePrefix = "arXiv",
    primaryClass = "gr-qc",
    doi = "10.1140/epjc/s10052-025-14785-8",
    journal = "Eur. Phys. J. C",
    volume = "85",
    number = "9",
    pages = "1043",
    year = "2025"
}

@article{Brassel:2022mss,
    author = "Brassel, Byron P. and Goswami, Rituparno and Maharaj, Sunil D.",
    title = "{Cosmic censorship and charged radiation in second order Lovelock gravity}",
    doi = "10.1016/j.aop.2022.169138",
    journal = "Annals Phys.",
    volume = "446",
    pages = "169138",
    year = "2022"
}

@article{Chatterjee:2021zre,
    author = "Chatterjee, Ayan and Ghosh, Avirup and Jaryal, Suresh C.",
    title = "{Gravitational collapse in Einstein-Gauss-Bonnet gravity}",
    eprint = "2108.11680",
    archivePrefix = "arXiv",
    primaryClass = "gr-qc",
    reportNumber = "USTC-ICTS/PCFT-21-35, 106, no.4, 044049",
    doi = "10.1103/PhysRevD.106.044049",
    journal = "Phys. Rev. D",
    volume = "106",
    number = "4",
    pages = "044049",
    year = "2022"
}

@article{Dadhich:2016wtb,
    author = "Dadhich, Naresh and Hansraj, Sudan and Chilambwe, Brian",
    title = "{Compact objects in pure Lovelock theory}",
    eprint = "1607.07095",
    archivePrefix = "arXiv",
    primaryClass = "gr-qc",
    doi = "10.1142/S0218271817500560",
    journal = "Int. J. Mod. Phys. D",
    volume = "26",
    number = "06",
    pages = "1750056",
    year = "2016"
}

\end{document}